\documentclass[final,3p,times]{elsarticle}

\usepackage{graphicx}
\usepackage{amssymb}

\usepackage{lineno}
\usepackage{graphicx}
\usepackage{float}
\usepackage{longtable,multicol,relsize }
\usepackage{algorithm,algcompatible,epsfig,amsfonts,amsmath}
\usepackage{bm}
\usepackage{tablefootnote}
\usepackage{multirow}
\usepackage{setspace}

\usepackage{caption}
\usepackage{subcaption}
\usepackage{amsmath}
\usepackage{wasysym}
\usepackage{bigints}
\usepackage{animate}

\usepackage{lineno}
\usepackage{epsfig,amsfonts,amsmath}
\usepackage{wrapfig}
\usepackage{breqn,soul}
\usepackage[colorinlistoftodos,prependcaption,textsize=tiny]{todonotes}

\def\correspondingauthor{\footnote{Corresponding author.}}

\newcommand{\vect}[1]{\boldsymbol{\mathit{#1}}}

\newcommand{\tens}[1]{\mathbf{#1}}

\newcommand{\tenf}[1]{{\mathbb{#1}}}%

\journal{}
\begin{document}

\begin{frontmatter}


\title{ A Knowledge-driven Physics-Informed Neural Network model; Pyrolysis and Ablation of Polymers}

\author[label1,label2]{Aref Ghaderi}
\author[label1,label4]{Ramin Akbari}
\author[label1,label3]{Yang Chen}
\author[label1,label5]{Roozbeh Dargazany\correspondingauthor{}}

\address[label1]{Department of Civil and Environmental Engineering, Michigan State University}

\address[label2]{ghaderi1@msu.edu}
\address[label4]{akbarigh@msu.edu}
\address[label3]{chenya36@msu.edu}
\address[label5]{roozbeh@msu.edu}

\begin{abstract}

In aerospace applications, multiple safety regulations were introduced to address associated with  pyrolysis. Predictive modeling of pyrolysis is a challenging task since multiple thermo-chemo-mechanical laws need to be concurrently solved at each time step. So far,  classical modeling approaches were mostly focused on defining the basic chemical processes (pyrolysis and ignite) at micro-scale by decoupling them from thermal solution  at the micro-scale and then  validating them using meso-scale experimental results. The advent of Machine Learning (ML) and AI in recent years has provided an opportunity to construct quick surrogate ML models to replace high fidelity multi-physics models, which have a high computational cost and may not be applicable for high nonlinear equations.

This serves as the motivation for the introduction of innovative Physics informed neural networks (PINNs)  to simulate multiple stiff, and semi-stiff ODEs that govern  Pyrolysis and Ablation.  Our Engine is particularly developed to calculate the char formation and degree of burning in the course of pyrolysis of crosslinked polymeric systems.  A multi-task learning approach is hired to assure the best fitting to the training data.
The proposed Hybrid-PINN (H-PINN) solver was bench-marked against finite element high fidelity solutions on different examples. We developed PINN architectures using  collocation training to forecast temperature distributions and the degree of burning in the course of  pyrolysis in multiple one- and two-dimensional examples. 
By decoupling thermal and mechanical equations, we can predict the loss of performance in the system by predicting the char formation pattern and localized degree of burning at each continuum.

\end{abstract}


\end{frontmatter}

\section{Introduction}
\label{Intro}

Heat causes both physical and chemical changes in solid polymeric materials. Thermal decomposition and thermal degradation need to be distinguished clearly from one another. The process of considerable chemical species change brought on by heat is known as thermal decomposition. Thermal degradation is the loss of physical, mechanical, or electrical qualities as a result of the effect of heat or excessive temperature on  material, product, or assembly. Thermal decomposition is the main alteration in burning.

Pyrolysis is one of the many different types of chemical decomposition processes that take place at higher temperatures. It differs from other processes like combustion and hydrolysis in that it seldom requires the addition of additional reagents such as oxygen ($O_2$) or water (in hydrolysis). Pyrolysis results in solids (char), condensable liquids (tar), and non-condensing/permanent gases \cite{natali2016ablation, chronopoulos2013thermal, liao2022hybrid}. Pyrolysis is widely used in the chemical industry to make ethylene, various kinds of carbon, and other compounds from petroleum, coal, and even wood, as well as to make coal coke. It has also been utilized lately on an industrial scale to convert natural gas (mainly methane) into non-polluting hydrogen gas and non-polluting solid carbon char. Pyrolysis might be used to turn biomass into syngas and charcoal, waste plastics back into useful oil, or trash into securely disposable substances, among other things \cite{dimitrienko1997thermomechanical, kumar2019advances}.

The combined impacts of thermal, chemical and physical processes play a significant role in the pyrolysis problems in polymeric materials \cite{luo2007thermo, luo2011thermo}. Table \ref{table1} summarizes the different thermal, chemical, and physical processes that may be investigated in burning.

\begin{table}[H]
\centering
\caption{A summary of processes that can be anticipated in burning analysis.}
\label{table1}
\begin{tabular}{c c}
\hline
\begin{tabular}[c]{@{}c@{}}Heat conduction/temperature \end{tabular} & \cite{henderson1985model}, \cite{henderson1987mathematical}, \cite{florio1991study}  \\ 
\begin{tabular}[c]{@{}c@{}}Pyrolysis \end{tabular}  & \cite{sullivan1992finite}, \cite{sullivan1990finite}, \cite{mouritz2009review}  \\ 
\begin{tabular}[c]{@{}c@{}}Volatile convective flow \end{tabular}  & \cite{carpier2020tensile}, \cite{feih2012tensile}, \cite{shi2013high} \\
\begin{tabular}[c]{@{}c@{}}Char formation/mass loss \end{tabular}  & \cite{carpier2020tensile}, \cite{feih2012tensile}, \cite{shi2013high}  \\
\begin{tabular}[c]{@{}c@{}}Internal pressure \end{tabular}  & \cite{sullivan1992finite}, \cite{sullivan1990finite}, \cite{mouritz2009review}  \\
\begin{tabular}[c]{@{}c@{}}Thermal expansion \end{tabular}  & \cite{mouritz2009review}, \cite{feih2012tensile}, \cite{shi2013high}   \\
\begin{tabular}[c]{@{}c@{}}Thermal stresses \end{tabular}   & \cite{kulkarni2018current}, \cite{li2020simulation}, \cite{vatikiotis1980heat}  \\ \hline
\end{tabular}
\end{table}

Before getting into the details of the models, it is crucial to understand what happens when the polymer is exposed to a high-temperature heat source. 

\begin{figure}[H]
\centerline{\includegraphics[width=.8\textwidth]{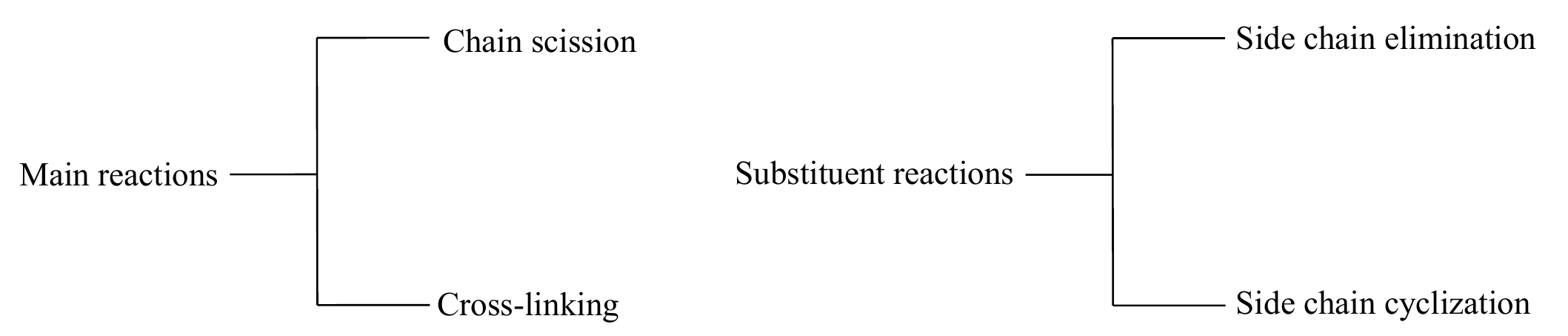}}
    \caption{General decomposition mechanisms.}
    \label{reactions}
\end{figure}

\textit{Chemical Processes:}
Polymers can decompose thermally by oxidative reactions or just by being exposed to heat. Several main groups of chemical processes play significant roles in the heat decomposition of polymers, including (1) Chain-stripping, in which atoms or groups not a part of the polymer chain (or backbone) are cleaved; (2) cross-linking, in which bonds are formed between polymer chains. (3) Random-chain scission, in which chain scissions occur at seemingly random locations in the polymer chain; (4) End-chain scission, in which individual monomer units are successively removed at the chain end; It suffices to state here that heat decomposition of a polymer often involves many reactions from each of these kinds. However, these broad categories offer a conceptual framework that is helpful for comprehending and categorizing polymer decomposing behavior (see Fig. \ref{reactions}).

\textit{Physical Processes:}
The nature of the material may have an impact on the many physical processes that take place during heat decomposition. For instance, simple phase transitions upon heating are not achievable for thermosetting polymeric materials because they are infusible and insoluble after they have been synthesized. On the other hand, if heating does not go over the minimum thermal breakdown temperature, thermoplastics can be softened by heating without the material undergoing irreversible alterations. This gives thermoplastic materials a significant edge in terms of how simple it is to mold or thermoform items. Carbonaceous chars are created during the thermal decomposition of several materials, including cellulosic, thermosetting, and thermoplastic ones. The ongoing thermal decomposition process will be significantly impacted by the physical makeup of these chars. The pace of heat decomposition of the remaining polymer is frequently determined by the physical properties of the char. Even though char production is a chemical process, its relevance is mostly because of its physical characteristics.

The polymer continues to decompose endothermically until the reaction zone reaches the material's back-face when the rest of the polymer is degraded to volatiles and char.

\begin{figure}[H]
\centerline{\includegraphics[width=.55\textwidth]{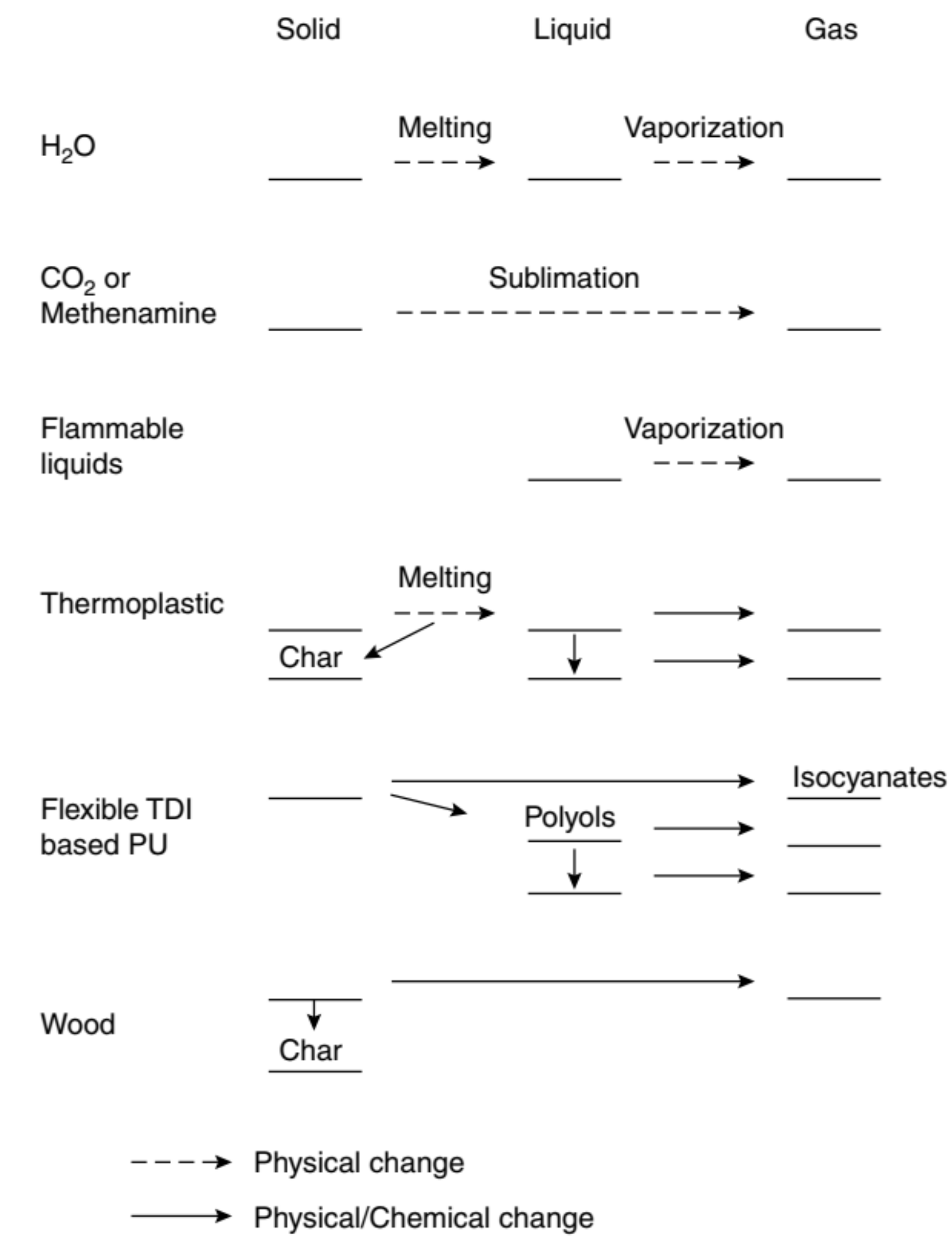}}
    \caption{Physical and chemical changes during thermal decomposition \cite{beyler2002thermal}.}
    \label{process}
\end{figure}


The physics of this thermo-chemical reaction is well understood, and it is described by a collection of coupled nonlinear partial differential equations (PDEs) that describe heat conduction and burning kinetics \cite{johnston1996process, boyard2016heat}. These PDEs (see section \ref{heat}), however, do not have a closed-form solution; computational approximation is necessary.

\textbf{\textit{Traditional solvers vs. Data-driven methods.}} The finite element method (FEM) is a common approach that (1) uses a set of basis functions within discretized sub-domains of the geometry to approximate the solution to the PDE at each time point and (2) iteratively simulates the evolution of the basis function coefficients forward in time \cite{zienkiewicz2005finite, boisse2005mesoscopic}. Comes with a restrictive computational cost in the case of repetitive simulations, such as optimization, control, real-time monitoring, probabilistic modeling, and uncertainty quantification. This is due to the fact that forward simulations require solving large non-linear equations repeatedly.

In general engineering applications, the advent of Machine Learning (ML) and AI in recent years has provided an opportunity to construct quick surrogate ML models to replace classical FEM. Classic neural networks, on the other hand, map across finite-dimensional spaces and can thus only learn discretization-specific solutions. This is frequently a constraint in actual applications, necessitating the creation of mesh-invariant neural networks. The finite-dimensional operators and Neural-FEM are two popular neural network-based techniques for solving multiple PDEs \cite{kochkov2021machine} 

\textbf{\textit{Finite-dimensional operators.}} 
A new line of research has suggested using neural networks to learn mesh-free, infinite-dimensional operators \cite{hornik1989multilayer, khoo2021solving}. Only a forward pass of the network is required to provide a solution for a new instance of the parameter, avoiding the significant computational difficulties that plague Neural-FEM approaches. The neural operator just requires data, not knowledge of the underlying PDE. Due to the difficulty of evaluating integral operators, neural operators have not provided effective numerical techniques that can match the success of convolutional or recurrent neural networks in the finite-dimensional scenario.

\textbf{\textit{Neural-FEM.}} The third method is known as the physics-informed neural network (PINN). PINN differs from other machine learning paradigms widely employed for mechanics and physics challenges in terms of how data is needed and employed.  Unlike supervised learning, which is often used for materials laws and requires artificial intelligence to be trained with labels in order to generate forecasts, the search for the solution in PINN does not require any data other than the ones required to form the loss function \cite{bar2019unsupervised, karniadakis2021physics, chen2021physics, haghighat2021physics}. 

\textbf{\textit{PINN challenges.} }The training of PINN, however, is far from simple, especially for nonlinear systems of equations. To construct the multi-layer perceptron, non-linearities should be applied to each element of the output of the linear transformation. This is unlike the finite element method, which is a more entrenched framework with clear strategies and established mathematical analysis that guarantees convergence and stability for both the solution and weighting functions in predetermined finite-dimensional spaces. Furthermore, for both forward and inverse problems, the physical constraints or controlling equations could be expressed in several ways; for instance, the collocation-based loss function, which evaluates the solution at specific collocation points, or the energy-based method that can reduce the order of the derivatives in governing equations despite requiring numerical integrations. A large number of tunable hyperparameters, such as the configurations of the neural network, the types of activation functions, and the neuron weight initialization, as well as different techniques to impose boundary conditions while providing significant flexibility, may bewilder researchers who are unacquainted with neural networks \cite{bhattacharya2020model,  psaros2021meta, fuhg2021model}.

\textbf{\textit{Goals.}} The training element of the physics-informed neural network for high nonlinear PDEs like Burning is the core of this article. Our goal is to find techniques to make PINN training less costly, lower the amount of trial-and-error necessary to appropriately tune the hyper-parameters, and at the very least, empirically increase the training process's robustness. Although the strategies given in this research may be relevant to other versions, we confined the scope of this study to the collocation physics-informed neural network. Our current attempt is to present a synthesis that incorporates:
(i) employing non-dimensionalization and normalization of the physical parameter to address the problem of complex equations, 
(ii) infusing a broad understanding of the solution field with imprecise knowledge, 
(iii) using a weighted-sum scheme to enhance optimization algorithms in the context of multi-objective optimization, 
(IV) exploiting different approaches for weight initialization due to unbalanced gradients causes inaccuracy in optimization.


\section{Exothermic Heat Transfer}
\label{heat}
When charring polymers are subjected to high temperatures, thermal energy is transferred into the polymer via thermal convection. Pyrolytic gases and solid residues will arise from the material's decomposition. According to the degree of pyrolysis, the polymer may be divided into three zones, as indicated in Figure \ref{process}: The polymer decomposes into three zones: charring, pyrolysis, and virgin material. \cite{dennis2019combustion}.

In the thermal study of heat transfer in polymers, three main types of thermal energy transmission are often considered: conduction, convection, and radiation. However, for the sake of simplicity, all mathematical models for polymers cover the effect of heat conduction in the case of one-sided heating only. The effect of external convection on heat transfer is rarely explored. Similarly, heat radiation from a polymer is rarely taken into account. So, the following PDE governs heat transfer in polymers, i.e., heat transfer with internal heat generation is expressed as \cite{bogetti1992process, tadini2017thermal, niaki2021physics}

\begin{equation}
\label{pde2}
    \frac{\partial}{\partial t}\left(\rho C_{p} T\right)=\frac{\partial}{\partial x}\left(k_{x x} \frac{\partial T}{\partial x}\right)+\frac{\partial}{\partial y}\left(k_{y y} \frac{\partial T}{\partial y}\right)+\frac{\partial}{\partial z}\left(k_{z z} \frac{\partial T}{\partial z}\right)+\dot{Q},
\end{equation}
where $T$ is the temperature, $C_{p}(T, \alpha)$, $k(T, \alpha)$, and $\rho(T, \alpha)$ are the solid's specific heat capacity, conductivity, and density which are a function of temperature and degree of burning, and $\dot{Q}$ is the rate of internal heat consumption. The change in thermal energy per unit volume is represented on the left-hand side of the equation, while the energy flux owing to conduction is represented on the right-hand side. The through-thickness direction is defined by the x-direction, whereas the planar directions are defined by the y- and z-directions.

Internal heat consumption in polymeric material is expressed as a function of the degree of burning $\alpha \in (0,1)$, which is a measure of the conversion achieved during the polymeric material's burning reactions. The relationship between $\dot{Q}$ and $\alpha$, in particular, can be expressed as \cite{mouritz2006modelling, mouritz2004thermomechanical}

\begin{equation}
    \dot{Q}=-Q_r \frac{d \alpha}{d t},
\end{equation}
where $Q_r$ is the heat of reaction generated per unit mass of polymer during burning \cite{henderson1987mathematical}. The evolving degree of burning in polymeric materials is usually controlled by an ordinary differential equation that indicates the rate of burning as a function of immediate temperature  and degree of burning \cite{henderson1985model, gibson2003modelling, zobeiry2021physics}.

\begin{equation}
\label{eq2}
    \frac{d \alpha}{d t} = g(\alpha ,T ) > 0.
\end{equation}

\textit{Assumptions:} Energy transfer via convection is considered to be insignificant in this study, and the volatile gases created by the pyrolysis reaction are expected to be evacuated from the material promptly and hence have no effect on the temperature. Temperature and the stage of the breakdown reaction influence thermal conductivity, density, and heat capacity. However, because the change in thermal conductivity with temperature cannot be calculated theoretically, this term must be determined experimentally throughout the temperature range of interest.

\begin{figure}[H]
\centerline{\includegraphics[width=1\textwidth]{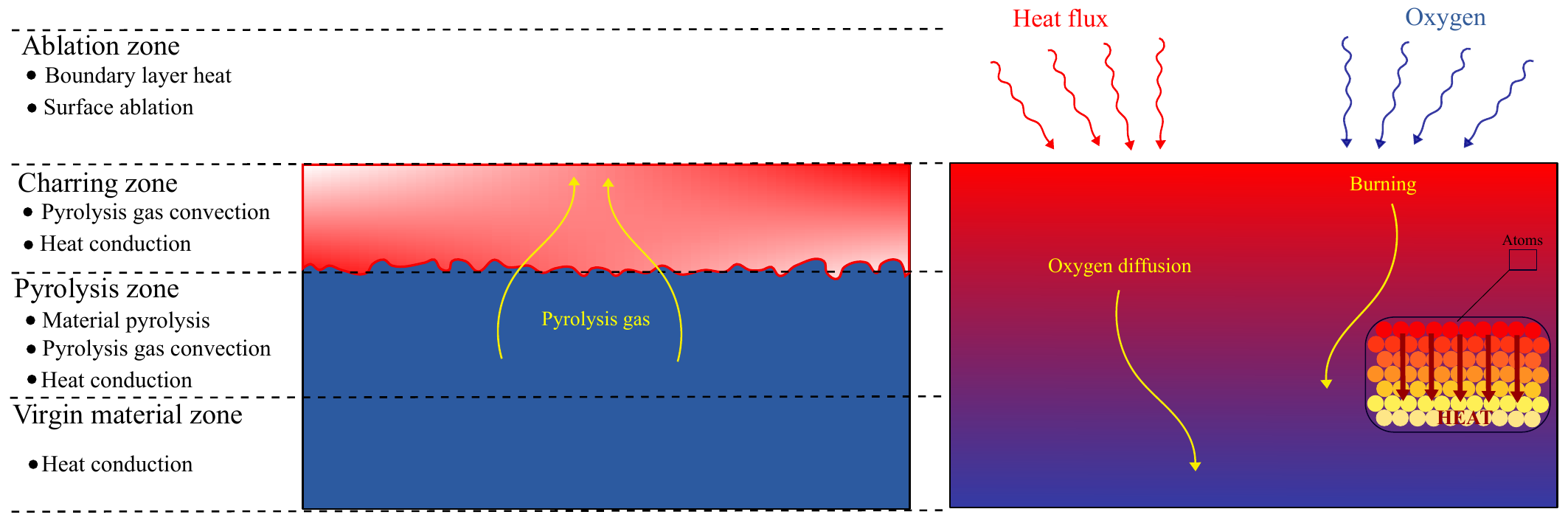}}
\caption{A schematic of polymeric materials during burning.}
\label{case}
\end{figure}

In the space-time domain $(\vect{x},t)$, this results in a coupled system of differential equations for temperature and degree of burning. The temperature $T(\vect{x},t)$ and degree of burning $\alpha (\vect{x},t)$ in the polymeric material are predicted by solving this system of differential equations with initial and boundary conditions. This work aims to solve the system of differential equations for the polymeric system burning shown in Fig. \ref{case}.

The boundary conditions at both ends can be set to a specific temperature by using the following equations

\begin{equation}
\small
\label{eq3}
\left.T\right|_{x=0, y, z}=T_{bx}(t), \qquad
\left.T\right|_{x=L, y, z}=T_{tx}(t), \qquad
\left.T\right|_{x, y=0, z}=T_{by}(t), \qquad
\left.T\right|_{x, y=L, z}=T_{ty}(t), \qquad
\left.T\right|_{x, y, z=0}=T_{bz}(t), \qquad
\left.T\right|_{x, y, z=L}=T_{tz}(t), \qquad
\end{equation}
where the subscripts $b$ and $t$ represent the bottom and top of materials, respectively. The initial conditions are written as follows

\begin{equation}
\label{eq4}
\left.T\right|_{t=0}=T_{0}(x) \qquad \text{and} \qquad
\left.\alpha \right|_{t=0}=\alpha_{0}(x),
\end{equation}
where $T_{0}$ is the system's initial temperature, which is normally assumed to be constant in the geometry, and $\alpha_{0}$ is the polymer's initial degree of burning, which is assumed to be zero for unburned materials in the entire spatial domain.

\section{Physics-Informed Neural Network}
\label{PINN}


\subsection{Neural Network Architecture}
Neural networks are well-known for their ability to represent information. Based on the universal approximation theorem, any continuous function can be arbitrarily estimated by a multi-layer perceptron containing one hidden layer and a finite number of neurons \cite{yarotsky2017error, berman2019survey}. While neural networks can compactly express very complicated functions, obtaining the precise parameters (weights and biases) required to solve a particular PDE can be challenging \cite{bin2021pinneik}. 

The bulk of solutions has used feed-forward neural networks since Raissi et al. \cite{raissi2019physics, haghighat2021sciann} original vanilla PINN. Some researchers, on the other hand, have tested with several types of neural networks to investigate their effect on the overall PINN performance. 

We start by building a simple $D$-layer multilayer feed-forward neural network comprising an input layer, $D -1$ hidden layers, and an output layer. We suppose that the $d^{th}$ hidden layer has $N_d$ neurons. The previous layer's post-activation output $\vect{x}^{d-1} \in \tenf{R}^{N_{d-1}}$ is then fed into the $d^{th}$ hidden layer, and the specific affine transformation is of the form

\begin{equation}
\mathcal{H}_{d}\left(\mathbf{x}^{d-1}\right) \triangleq \mathbf{W}^{d} \mathbf{x}^{d-1}+\mathbf{b}^{d},
\end{equation}
where the network weight $\mathbf{W}^{d} \in \tenf{R}^{N_{d} \times N_{d-1}}$ and the bias term $\mathbf{b}^{d} \in \tenf{R}^{N_{d}}$ to be learned are both initialized using unique procedures like Xavier or He initialization \cite{glorot2010understanding, he2015delving}.

The nonlinear activation function $\sigma(.)$ is applied component-by-component to the current layer's affine output $\mathcal{H}_{d}$. Furthermore, for some regression issues, this nonlinear activation is not employed in the output layer. As a result, the neural network may be denoted as

\begin{equation}
\mathcal{N}(\mathbf{x} ; \Theta)=\left(\mathcal{H}_{D} \circ \sigma \circ \mathcal{H}_{D-1} \circ \cdots \circ \sigma \circ \mathcal{H}_{1}\right)(\mathbf{x}),
\end{equation}
where $\circ$ denotes the composition operator, $\Theta=\left\{\mathbf{W}^{d}, \mathbf{b}^{d}\right\}_{d=1}^{D} \in \mathcal{P}$ denotes the learnable parameters to be optimized later in the network, and $\mathcal{P}$ denotes the parameter space, and $\mathcal{N}$ and $\mathbf{x}^0=\mathbf{x}$ denote the network's output and input, respectively.

 
\paragraph{\textbf{Activation function}}
DNN training performance is influenced by the activation function. Activations like as ReLU, Sigmoid, and Tanh are frequently utilized \cite{sun2020surrogate, ertuugrul2018novel}. 
Because the activation function in a PINN framework is evaluated using the second-order derivative, it is critical to choose it carefully. Because most activation functions (such as Sigmoid, Tanh, and Swish) are nonlinear around $0$, it is preferable to pick a range of [$0, 1$] rather than a wider domain when rescaling the PDE to a dimensionless form. Furthermore, smooth activation functions such as the sigmoid and hyperbolic tangent can be used to ensure the regularity of PINNs, allowing for accurate calculations of PINN generalization error\cite{mishra2021estimates, krishnapriyan2021characterizing}.

\subsection{Automatic Differentiation}
In order to solve a PDE in PINNs, it is necessary to take derivatives of the network's output with respect to the input. The function $\boldsymbol{u}$ can be differentiated since it is approximated by a NN with smooth activation function $\hat{\boldsymbol{u}}_{\Theta}$. Calculating derivatives can be done in four ways: manually, symbolically, numerically, or automatically. When applied to complex functions, symbolic and numerical methods such as finite differentiation perform poorly; on the other hand, automatic differentiation (AD) conquers many limitations such as floating-point precision errors, numerical differentiation, and memory-intensive symbolic approaches \cite{wang2022and, bolte2020mathematical}. 

\paragraph{\textbf{Nondimensionalization \& Normalization of Equations}}

Nondimensionalization $\bar X=\frac{X}{X_0}$ will be used to remove of physical dimensions of the governing parameters $X$ to describe the physical process as the deviation of the parameters from their reference point $X_0$ which is used for normalization. This procedure allows us to simplify the loss function and prevent the effects of measured units in dimensional analysis. Further scaling of the non-dimensionless parameters is also  used  to prevent the effects of massive fluctuation of   certain quantities  that are less important  relative to some appropriate unit. These units refer to quantities intrinsic to the physics of the process rather than measured units.   Consider the dimensionless variables (denoted by $\overline{\mathrm{o}}$) as follows:

\begin{equation}
\bar{t}=\frac{t}{t^{*}}, \quad \overline{{x}}=\frac{{x}}{x^{*}}, \quad \overline{{y}}=\frac{{y}}{y^{*}}, \quad \bar{T}=\frac{T}{T^{*}}, \quad \bar{\alpha}=\frac{\alpha}{\alpha^{*}},
\end{equation}
$t^{*}, x^{*}, y^{*}, T^{*}, \alpha^{*}$ are dimensionless variables that written as

\begin{equation}
x^{*} = {L_x}, \qquad y^{*} = {L_y}, \qquad t^{*} = \frac{\rho C_p {L_x^2  L_y^2}}{k_{xx} L_y^2 + k_{yy} L_x^2} , \qquad T^{*} = {T_t - T_0}, \qquad \alpha^{*} = 1.
\end{equation}

As a result, the partial derivatives and the divergence operator are likewise stated in this manner \cite{haq2020heat}.

\begin{equation}
\frac{\partial \circ}{\partial t}=\frac{1}{t^{*}} \frac{\partial \circ}{\partial \bar{t}}, \quad {\nabla} \cdot \circ=\frac{1}{x^{*}} \overline{{\nabla}} \cdot \circ, \quad
{\nabla} \cdot \circ=\frac{1}{y^{*}} \overline{{\nabla}} \cdot \circ.
\end{equation}

The degree of burning relations and the heat equation are written in a dimensionless form as

\begin{equation}
\label{pde1}
    \frac{1}{t^{*}} \frac{\partial }{\partial \bar{t}}\left(\rho C_{p} \bar{T}\right)=\frac{1}{x^{*}} \frac{\partial }{\partial \bar{x}}\left(k_{x x} \frac{1}{x^{*}} \frac{\partial \bar{T}}{\partial \bar{x}}\right)+\frac{1}{y^{*}} \frac{\partial }{\partial \bar{y}}\left(k_{y y} \frac{1}{y^{*}} \frac{\partial \bar{T}}{\partial \bar{y}}\right)+\frac{1}{z^{*}} \frac{\partial }{\partial \bar{z}}\left(k_{z z} \frac{1}{z^{*}} \frac{\partial \bar{T}}{\partial \bar{z}}\right)+\dot{Q}, \quad \textit{while} \quad\dot{Q}=-Q_r \frac{1}{t^{*}} \frac{\partial \bar{\alpha}}{\partial \bar{t}}.
\end{equation}

\subsection{Model Estimation by Learning}
In the PINN methodology, network training takes place by minimizing the total loss of the network parameters $\boldsymbol{\Theta}$,

\begin{equation}
\boldsymbol{\Theta}^{*}=\underset{\boldsymbol{\Theta} \in \mathbb{R}^{D}}{\operatorname{argmin}} \mathcal{L}_{\mathcal{T}}(\mathbf{X} ; \boldsymbol{\Theta}).
\end{equation}

An error or loss function is defined in PINNs utilizing the network's processed outputs and derivatives based on the equations guiding the problem's physics. As a result, the network's total loss, $\mathcal{L_T}$, is made up of the sum of loss terms for the PDE ($\mathcal{L_{\mathcal{F}}}$) and the initial and boundary conditions ($\mathcal{L_{\mathcal{B}}}$). Also, let us assume that we have some imprecise knowledge which can provide a general idea of the solution field. By updating the loss function, the imprecise knowledge can be easily included in the total loss function to guide the training process as

\begin{equation}
    \mathcal{L_T}(\Theta)=\mathcal{L_{\mathcal{F}}} (\Theta) + \mathcal{L_{\mathcal{B}}} (\Theta)+ \sum_{i=1}^{N_{back}} \eta_i \mathcal{L}_i,
\end{equation}
where $\mathcal{L}_i$ is the loss term for the $i_{th}$ background knowledge, and $\eta_i$ is the control term for each knowledge's contribution. Since the imprecise knowledge is not exact, the minimizer is far away from any true optimal point. As a result, we are interested in an adaptive optimization problem in which imprecise knowledge is originally integrated into a supervised learning assignment. Setting $\eta_i = 0$ turns off the associated objective once the solution is close enough to finish the supervised learning work.. $\mathcal{L_{\mathcal{F}}}$ represents due to error in satisfying the governing differential equations $\mathcal{F}$. It imposes the differential equation $\mathcal{F}$ at collocation points over the domain, $\Omega$, which can be chosen uniformly or unevenly. The other term, $\mathcal{B}\left(\hat{\boldsymbol{u}}_{\Theta}\right)=\boldsymbol{g}$, represents due to the error in satisfying of the boundary or initial conditions. A mean square error formulation is used in a common implementation of the loss, where

\begin{equation}
\mathcal{L}_{\mathcal{F}}(\Theta)={{M S E}}_{\mathcal{F}}=\frac{1}{N_{c}} \sum_{i=1}^{N_{c}} \| \mathcal{F}\left(\hat{\boldsymbol{u}}_{\Theta}\left(\boldsymbol{z}_{i}\right)\right)-\boldsymbol{f}\left(\boldsymbol{z}_{i}\right))\|^{2}=\frac{1}{N_{c}} \sum_{i=1}^{N_{c}}\| r_{\Theta}\left(\boldsymbol{u}_{i}\right)-\boldsymbol{r}_{i} \|^{2},
\end{equation}
and 

\begin{equation}
\mathcal{L}_{\mathcal{B}}(\Theta)=M S E_{\mathcal{B}}=\frac{1}{N_{B}} \sum_{i=1}^{N_{B}} \| \mathcal{B}\left(\hat{\boldsymbol{u}}_{\Theta}(\boldsymbol{z})\right)-\boldsymbol{g}\left(\boldsymbol{z}_{i}\right)) \|^{2}.
\end{equation}

\begin{figure}[H]
\centerline{\includegraphics[width=1\textwidth]{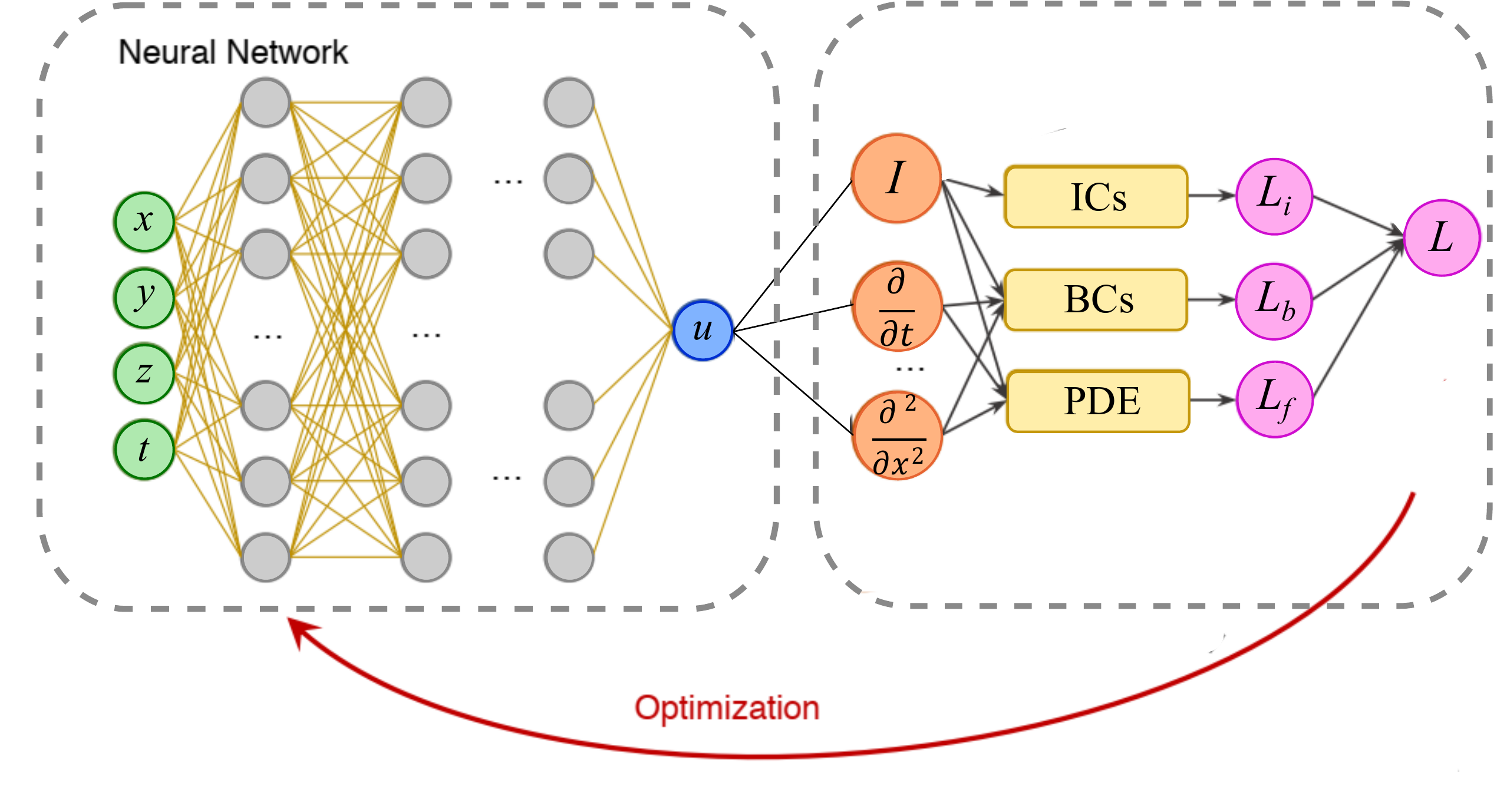}}
    \caption{A schematic of PINN blocks. Differential equation residual (loss) terms, as well as initial and boundary conditions, make up PINNs.}
    \label{blocks}
\end{figure}



Optimization is used to minimize the loss function. Based on the literature, optimization of loss functions is performed using minibatch sampling with the Adam and LBFGS method, which is a quasi-Newton optimization procedure in most of the PINN literature. 
The gradient-based optimizer will almost likely become stuck in one of the local minima for the loss function \cite{pang2019fpinns, maclaurin2015gradient}. Because stochastic gradient descent (SGD) struggles with random collocation points, especially in 3D setups, the Adam approach, which combines adaptive learning rate and momentum approaches, is used to speed up convergence \cite{zhu2021machine, bock2019proof}.

Note that because it may incorporate numerous different objectives for separate tasks, the PINN problem might be regarded as a multi-objective multi-task problem. This view of modeling has several advantages. It lowers the order of regularity, which is advantageous for problems involving discontinuities. Additionally, because the order of partial derivatives increases exponentially in terms of time complexity in the automated differentiation technique, it may lower the total computing cost \cite{zhu2021local}. Multiple objectives used to assess the correctness of the solution may result in gradients that are conflicting. As a result of the opposing gradient, a compromised solution that is not always Pareto optimal may emerge. We use the gradient surgery technique \cite{yu2020gradient} for the PINN framework to address the conflicting gradient and the challenge of balancing different objectives. We have the following update rule in the most common form of GD-based methods,

\begin{equation}
    \Theta_n = \Theta_{n-1} - \alpha_n \nabla_{\Theta} {\mathcal{L}_{\mathcal{T}}}_{n-1},
\end{equation}
where $\Theta_n$ is the updated unknown vector at iteration $n^{th}$, $\nabla_{\Theta} {\mathcal{L}_{\mathcal{T}}}_{n-1}$ is the gradient of the total loss function with respect to the unknown vector at iteration $n-1$, and $\alpha_n$ is the learning rate that regulates the solution's stability during iterations.

\begin{figure}[H]
\centerline{\includegraphics[width=.9\textwidth]{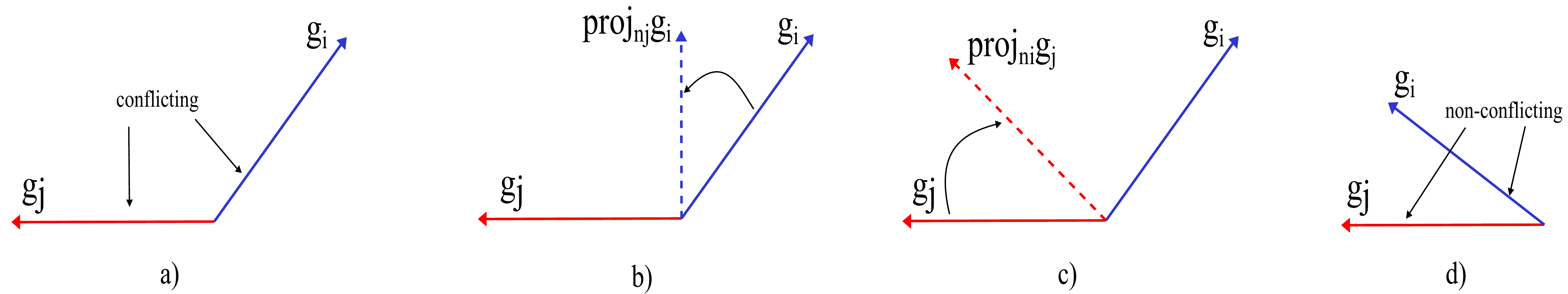}}
\caption{Gradient surgery idea \cite{yu2020gradient, bahmani2021training}}
\label{surgery}
\end{figure}

Let's call each objective's gradient vector $\vect{g}_i = \nabla_{\Theta} {\mathcal{L}_{i}}(\Theta)$ in gradient surgery, and the total gradient is $\vect{g} = \nabla_{\Theta} {\mathcal{L}_{T}}(\Theta) = \sum_{i=1}^{N_{obj}} \vect{g}_i$. You can find the concept in Fig. \ref{surgery}. For further information, we direct interested readers to \cite{yu2020gradient}.

\begin{algorithm}[H]
    \caption{Algorithm of gradient surgery}
  \begin{algorithmic}[1]
    \STATE $\tens{\Theta}$  $\leftarrow$ Initialize random weights and biases.
     \STATE $n, i$  $\leftarrow 1$ 
    \FOR{$n$ to $N_{itr}$}
        \FOR{$i$ to $N_{obj}$}
    \STATE $\mathcal{L}_i (\tens{\Theta_n})$, $\vect{g_i}=\nabla \mathcal{L}_i $ $\leftarrow$ compute objectives and gradients.
            \IF{$\vect{g_i} . \vect{g_j} < 0$ }
    \STATE $\vect{g_i}=\vect{g_i}-\frac{\vect{g_i}.\vect{g_j}}{||\vect{g_i}||^2} $ $\leftarrow$ modified gradient.
            \ENDIF
            \STATE $\vect{g} = \sum_{i=1}^{N_{obj}} \vect{g}_i$ $\leftarrow$ Aggregate gradient vectors.
        \ENDFOR
        \STATE $\tens{\Theta}_{n+1} $ $\leftarrow$ $GD(\tens{\Theta}_{n}, \vect{g})$
    \ENDFOR
  \end{algorithmic}
\end{algorithm}

\paragraph{\textbf{Weight initialization}} Because training deep models is such a challenging operation, the choice of initialization has a significant impact on most algorithms. The beginning point can decide whether or not the method converges at all, with certain initial points being so insecure that the algorithm runs into arithmetic issues and fails. Over the previous decade, more customized techniques have been the de facto norm, as they may result in a slightly more successful optimization process. Pang et al. \cite{pang2019fpinns} describe a method for picking the most appropriate one. 

For initialization, Glorot and Bengio \cite{glorot2010understanding} advocated using a correctly scaled uniform distribution. This is referred to as the "Xavier" initialization. The assumption that the activations are linear is used to derive it. This approach is computed as a random number with a uniform probability distribution ($\mathcal{U}$) as "$weight=\mathcal{U}[-\frac{1}{\sqrt{n}}, \frac{1}{\sqrt{n}}]$", where $n$ is the number of node inputs. When used to initialize networks that employ the rectified linear (ReLU) activation function, the "Xavier" weight initialization was discovered to have issues.

The "He" initialization method is the current standard for initializing the weights of neural network layers and nodes that employ the rectified linear (ReLU) activation function \cite{he2015delving}. The he initialization technique is computed as a random number with a Gaussian probability distribution ($\mathcal{G}$) as $weight = \mathcal{G}(0,\sqrt{\frac{2}{n}})$, where $n$ is the number of node inputs.

\paragraph{\textbf{Adaptive weights}}
Training multi-objective total loss functions pose challenges for optimization techniques, as others have pointed out \cite{raissi2019physics}. The unique solution to the governing equations is obtained once initial and boundary conditions are imposed strongly, especially in the case of boundary value problems.  It is common to use a weighted-sum scheme to enhance optimization algorithms in the context of multi-objective optimization \cite{marler2010weighted}. Wang et al. \cite{wang2020understanding} showed that unbalanced gradients cause inaccuracy in optimization and developed an adaptive loss weight approach where gradients of individual terms in the loss function are normalized in order to decrease the stiffness of the gradient dynamics in PINNs.  As a result, the total loss term $\mathcal{L_{Tot}}$ is changed to

\begin{equation}
    \mathcal{L_{Tot}}= \omega_{\mathcal{F}} \mathcal{L_{\mathcal{F}}} + \omega_{\mathcal{B}} \mathcal{L_{\mathcal{B}}},
\end{equation}
where $\omega$ represents the weight loss associated with each loss term. Based on the parameter values $(\Theta)^{e}$ at epoch $e$, the updated scaling weight $\omega^{e+1}$ for each loss term is calculated as

\begin{equation}
\omega^{e+1}=\beta \omega^{e}+(1-\beta) \hat{\omega}^{e+1},
\end{equation}
where $\beta = 0.9$ based on \cite{wang2020understanding} and  

\begin{equation}
\hat{\omega}_{i}^{e+1}=\frac{1}{\omega_{i}^{e}} \frac{\max \left(\left|\nabla \mathcal{L}_{T}\left(\boldsymbol{\Theta}^{e}\right)\right|\right)}{\operatorname{mean}\left(\left|\nabla \mathcal{L}_{i}\left(\boldsymbol{\Theta}^{e}\right)\right|\right)}
\end{equation}

Such an adaptive strategy for loss weights improves the system's robustness and improves the model's prediction accuracy.

\begin{algorithm}[H]
    \caption{PINN Algorithm for Burning Problem}
  \begin{algorithmic}[1]
    \STATE $\textbf{X}, \textbf{t}$  $\leftarrow$ Sample uniformly spatial and temporal domains.
    \STATE $\tens{\Theta_n}$ $\leftarrow$ Initialize randomly using Xavier scheme.
    \STATE $n$  $\leftarrow 1$ 
    \WHILE{$err > TOL$}
    \STATE $\tens{\Theta_n}$ $\leftarrow$ Optimize $\mathcal{L}_T$ for $\tens{\Theta}$.
    \STATE $T_n, \alpha_n$ $\leftarrow$ Evaluate temperature and degree of burning using $\tens{\Theta}$.
    \STATE $err$ $\leftarrow$ $\left\|\tens{{\Theta}_{n}}-\tens{{\Theta}_{n-1}}\right\| /\left\|\tens{{\Theta}_{n}}\right\|$ with $\left\| o \right\|$ as the $L^2$ norm.
    \STATE $n$  $\leftarrow $ $n+1$ 
    \ENDWHILE
  \end{algorithmic}
\end{algorithm}

\section{Case Study 1:  1D Burning Problem of Polymer }
\label{resultf}
To evaluate the performance of the proposed PINN, the heat transfer PDEs in a polymeric material are trained in Python (V3.6.8), using Tensorflow and Keras libraries (V2.10). We can define the PINN solution approach for the 1D pyrolysis problem now that we have covered the method of Physics-Informed Neural Networks (PINNs) and heat equations.

\begin{equation}
\label{pde4}
    \frac{1}{t^{*}} \frac{\partial }{\partial \bar{t}}\left(\rho C_{p} \bar{T}\right)=\frac{1}{x^{*}} \frac{\partial }{\partial \bar{x}}\left(k_{x x} \frac{1}{x^{*}} \frac{\partial \bar{T}}{\partial \bar{x}}\right)+\dot{Q}, \quad \textit{while} \quad\dot{Q}=-Q_r \frac{1}{t^{*}} \frac{\partial \bar{\alpha}}{\partial \bar{t}}.
\end{equation}

$T$ and $\alpha$ are the unknown solution variables in the case of 1D burning. For the dimensionless version of these variables, neural networks are,

\begin{equation}
\bar{T}:(\bar{x}, \bar{t}) \mapsto \mathcal{N}_{\bar{T}}\left(\bar{x}, \bar{t} ; \boldsymbol{\Theta}_{\bar{T}}\right), \qquad \textit{and} \qquad
\bar{\alpha}:(\bar{x}, \bar{t}) \mapsto \mathcal{N}_{\bar{\alpha}}\left(\bar{x}, \bar{t} ; \boldsymbol{\Theta}_{\bar{\alpha}}\right),
\end{equation}
where $\boldsymbol{\Theta}_{o}$ denotes that these networks have their own set of parameters. After that, the total coupled loss function is written as,

\begin{equation}
    \mathcal{L}_{Tot}= \omega_{1} \mathcal{L}_{{T}} + \omega_{2} \mathcal{L}_{\alpha} + \omega_{3} \mathcal{L}_{\alpha_0} + \omega_{4} \mathcal{L}_{T_0} + \omega_{5} \mathcal{L}_{T_{BC1}} + \omega_{6} \mathcal{L}_{T_{BC2}}+ \sum_{i=1}^{N_{back}} \eta_i \mathcal{L}_i,
\end{equation}
while

\begin{equation}
\begin{aligned}
\mathcal{L}_{T} &=\frac{1}{N_{c}} \sum_{i=1}^{N_{c}} \| \frac{1}{t^{*}} \frac{\partial }{\partial \bar{t}}\left(\rho C_{p} \bar{T}\right)-\frac{1}{x^{*}} \frac{\partial }{\partial \bar{x}}\left(k_{x x} \frac{1}{x^{*}} \frac{\partial \bar{T}}{\partial \bar{x}}\right) + Q_r \frac{1}{t^{*}} \frac{\partial \bar{\alpha}}{\partial \bar{t}} \|^{2}, \\
\mathcal{L}_{\alpha} &=\frac{1}{N_{c}} \sum_{i=1}^{N_{c}} \| \frac{1}{t^{*}} \frac{\partial \bar{\alpha}}{\partial \bar{t}} - g(\bar{\alpha} ,\bar{T} ) \|^{2}, \qquad \textit{loss of burning degree} \\
\mathcal{L}_{T_{0}} &=\frac{1}{N_{B}} \sum_{i=1}^{N_{B}} \| \bar{T}-\bar{T}_{0}(x) \|^{2}, \qquad \textit{loss of initial condition of temperature} \\
\mathcal{L}_{T_{b c_{1}}} &=\frac{1}{N_{B}} \sum_{i=1}^{N_{B}} \| \bar{T}-\bar{T}_{b}(t) \|^{2},  \qquad \textit{loss of boundary condition at bottom}\\
\mathcal{L}_{T_{b c_{2}}} &=\frac{1}{N_{B}} \sum_{i=1}^{N_{B}} \| \bar{T}-\bar{T}_{t}(t) \|^{2}, \qquad \textit{loss of boundary condition at top} \\
\mathcal{L}_{\alpha_{0}} &=\frac{1}{N_{B}} \sum_{i=1}^{N_{B}} \| \bar{\alpha}-\bar{\alpha}_{0}(x) \|^{2} \qquad \textit{loss of initial condition of burning degree},
\end{aligned}
\end{equation}

\begin{figure}[H]
\centerline{\includegraphics[width=1\textwidth]{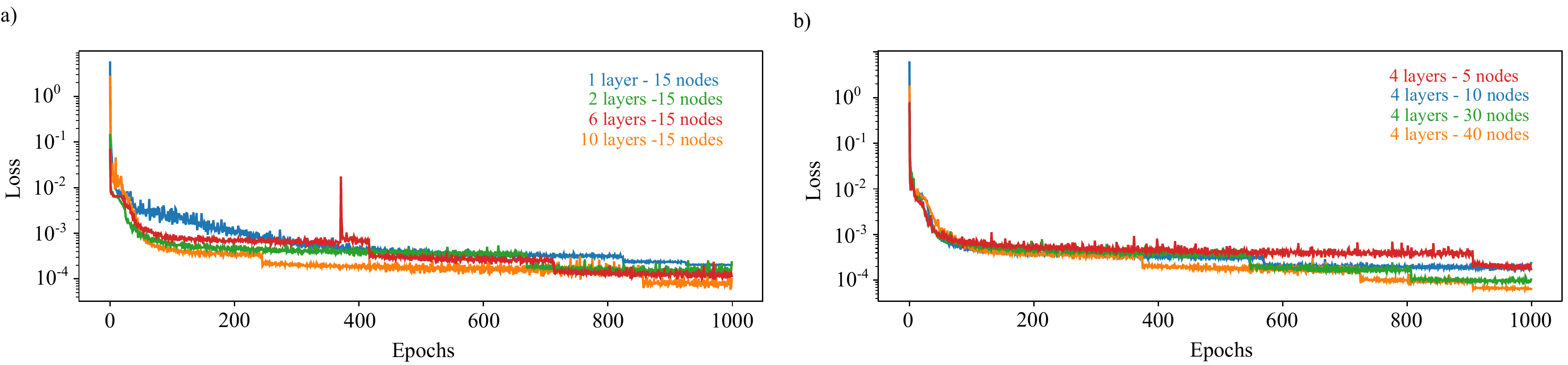}}
\caption{Losses for a variety of hidden layer (left) and neuron numbers (right).}
\label{loss sensitivity}
\end{figure}

\paragraph{\textbf{Training and Hyperparameter Searches}}
It is hardly surprising that network size, architecture, and optimizer hyperparameters like learning rate can significantly impact PINN solution quality. We used the hyperbolic-tangent and sigmoid activation functions to create neural networks with four hidden layers and 20 neurons in each layer for all of the examples in this paper. We used the Adam optimizer with an initial learning rate of $2*10^{-3}$ and an exponential learning decay to $10^{-5}$ at the end of training. We have also explored a range of hyperparameters, but we have not noticed any substantial improvements in terms of improving the pyrolysis problem studied here. Fig. \ref{loss sensitivity} shows the training history for various hidden layers and neurons.

\begin{figure}[H]
\centerline{\includegraphics[width=1\textwidth]{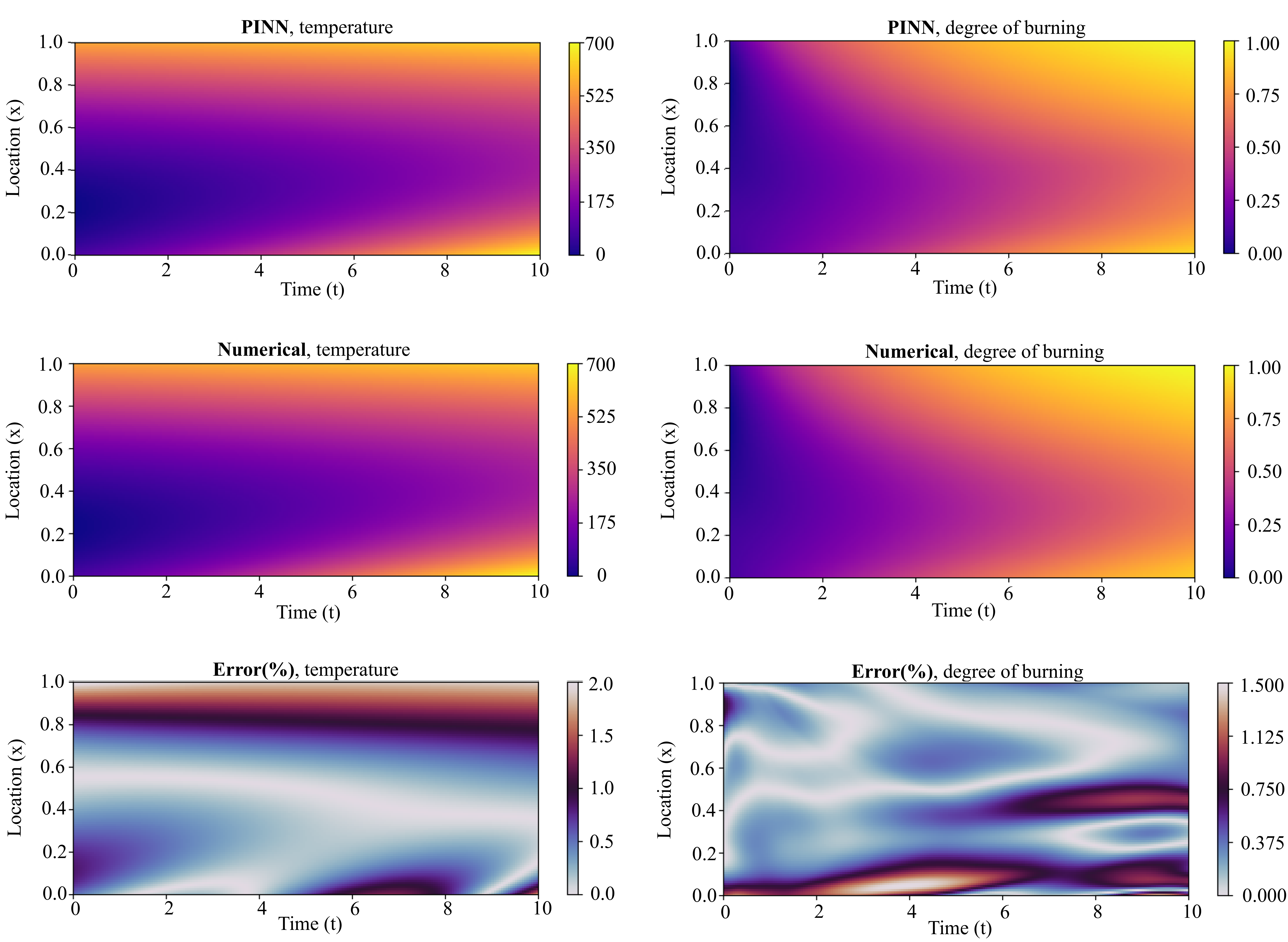}}
\caption{PINN and numerical predictions for full cycle of burning for temperature (left), and degree of burning (right).}
\label{1D}
\end{figure}

\paragraph{\textbf{Comparison with Numerical Results}}
The reference numerical model results will be compared to the above-mentioned best-performing PINN model outputs in this part. Fig. \ref{1D} shows the PINN, numerical solution, and errors. As you can see, the PINN solution has a high level of agreement with the expected solution when compared to the findings of numerical analysis. The upper face is given a temperature of $T_{t} = 700$, while the bottom face is given a temperature of $T_{t} = 700sin(\frac{\pi}{20} t)$. $T_{0} = 700x^3$ is the initial temperature.

\begin{figure}[H]
\centerline{\includegraphics[width=0.6\textwidth]{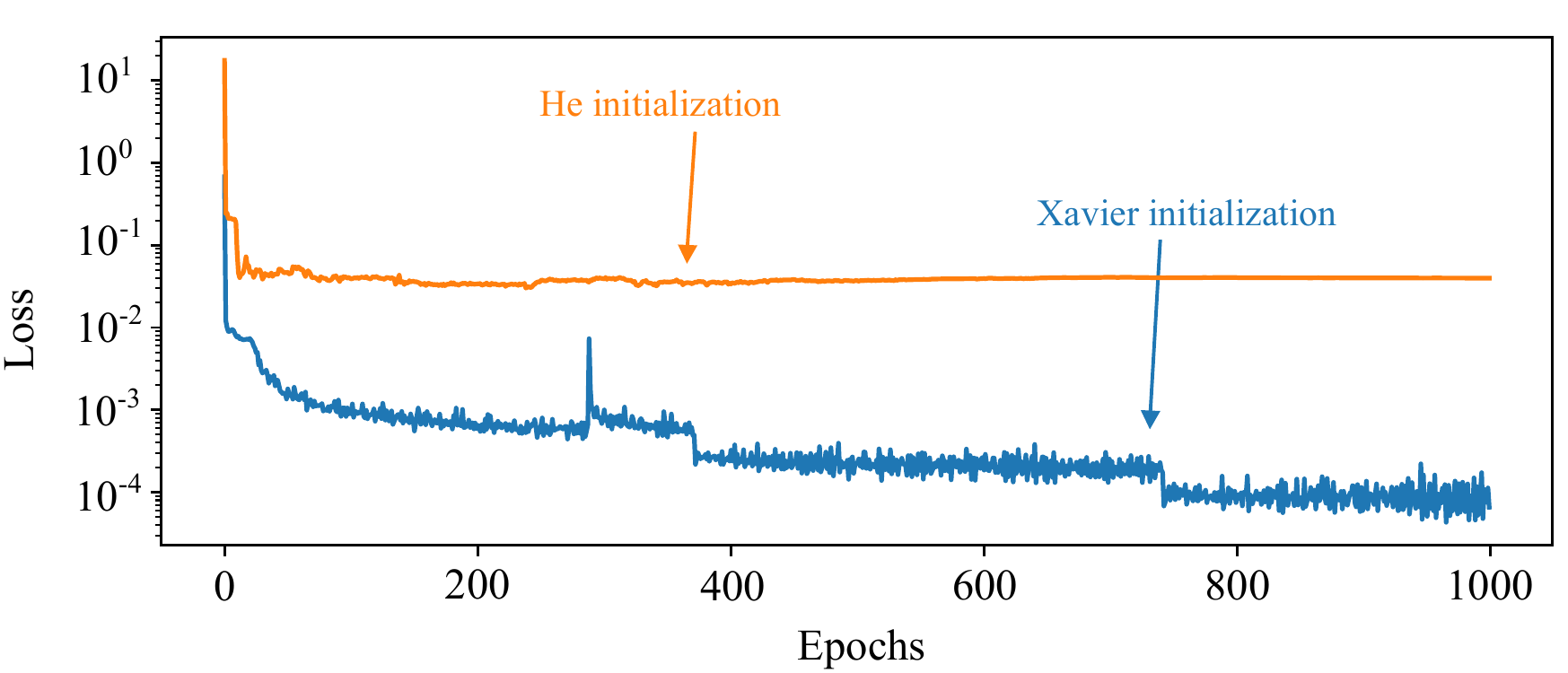}}
\caption{Evolution of loss term for Xavier weight initialization and He weight initialization.}
\label{weighttrain}
\end{figure}

\paragraph{\textbf{Role of Weight Initialization}} Because training deep models is such a tough operation, the choice of initialization has a significant impact on most algorithms. So, we trained the model with Xavier and He initialization for tanh and ReLU activation functions hereunder. Fig. \ref{weighttrain} shows the results for Xavier and He initialization.


\begin{figure}[H]
\centerline{\includegraphics[width=0.6\textwidth]{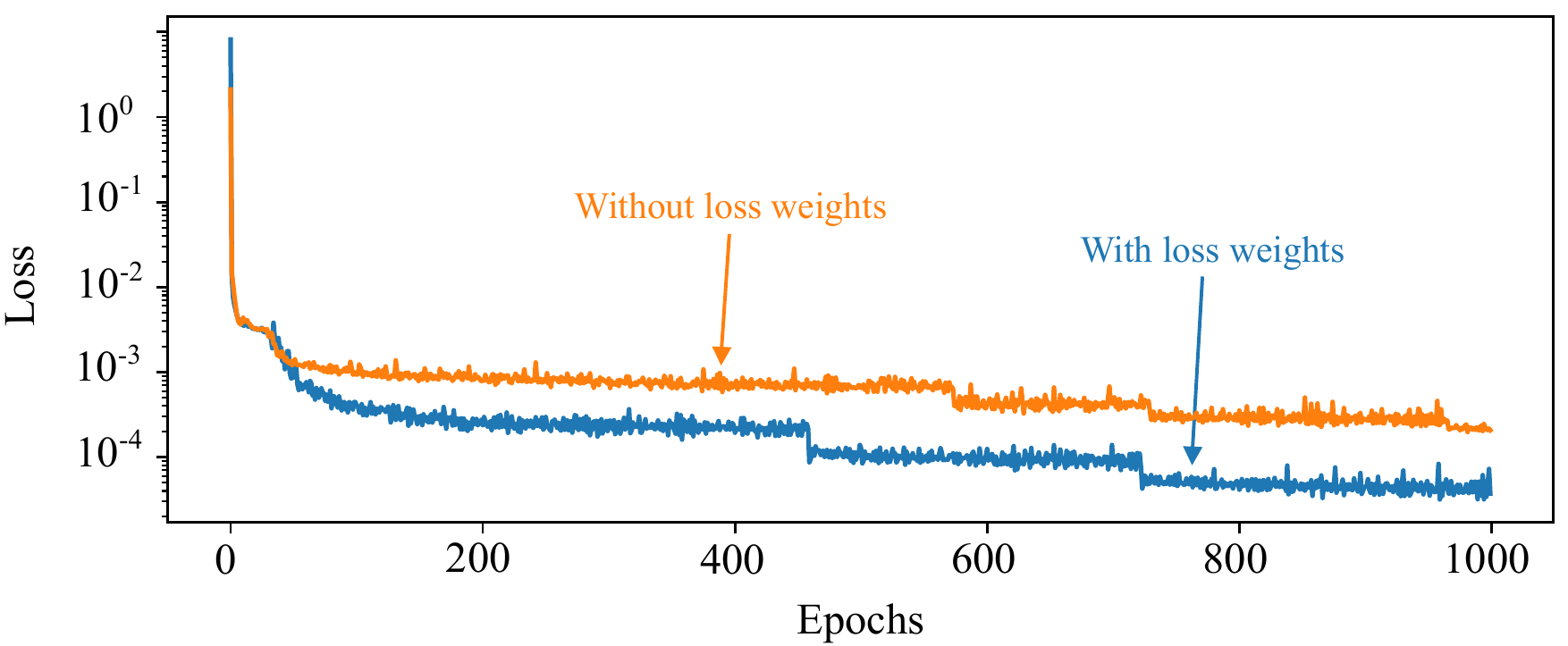}}
\caption{Evolution of loss term for optimization with and without loss weights.}
\label{lossweight}
\end{figure}

{\textit{Training with and without Adaptive Weights.}} The performance of the suggested PINN with and without loss weights is shown in Fig. \ref{lossweight} to emphasize the need to use loss weights.

\begin{figure}[H]
\centerline{\includegraphics[width=0.6\textwidth]{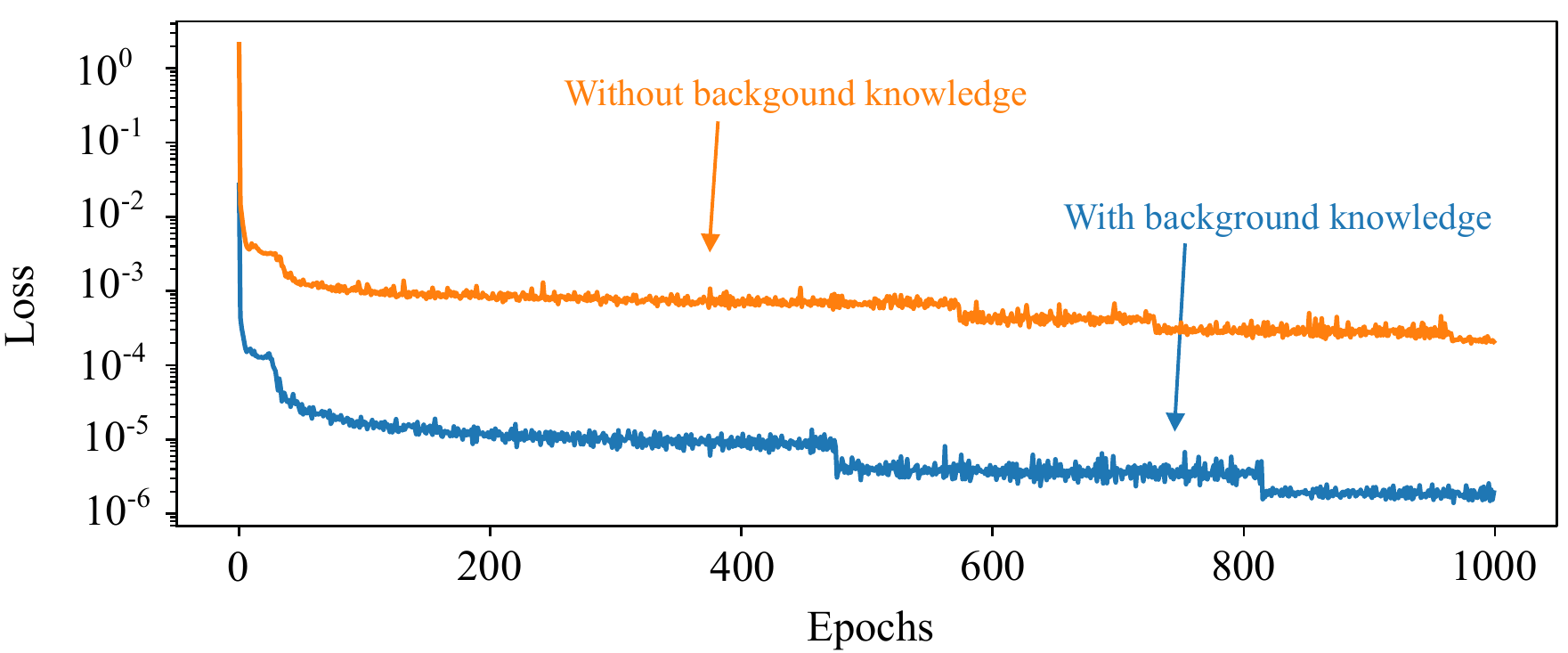}}
\caption{Evolution of loss term for optimization with and without background knowledge.}
\label{lossknow}
\end{figure}

{\textit{Training with and without Background Knowledge.}} To show the ability of background knowledge, we compare two case studies with and without background knowledge. We consider that the top of the geometry of polymer has a higher temperature and degree of burning always compared to the bottom of that. We considered it as the background knowledge and added it to the total loss. Fig. \ref{lossknow} shows the result for comparing of this factor.

\section{Case Study 2:  2D Burning Problem of Matrix Cube}
\label{2D}
For the second example, we modeled the pyrolysis of ethylene propylene diene monomer (EPDM) composites at the high temperature and pressure of solid rocket motors (SRMs), which results in a char layer with a non-uniform distribution pattern. Interestingly, the compact structure of the char can increase the ablation resistance of the composites.
To better understand the ablation mechanism and inform the design of EPDM composites for SRMs, it is crucial to understand and describe the char formation pattern throughout the life of the material and also understand the material behaviour at the interface.
We used our H-PINN  to predict the spatial evolution of chemical processes (pyrolysis and charring) before validating them against mesoscale finite element results.  It is of great theoretical and practical significance to study the formation mechanism of the compact structure of the char layer for further studying the ablation mechanism and guiding the development of EPDM composites. To date, the char formation pattern in polymeric materials cannot be predicted with reduced order models, such as machine-learned approaches. In this work, the hybrid PINN for 2D pyrolysis of  EPDM composites are designed  from the basic 2D   heat equation, namely

\begin{equation}
\label{pde3}
    \frac{1}{t^{*}} \frac{\partial }{\partial \bar{t}}\left(\rho C_{p} \bar{T}\right)=\frac{1}{x^{*}} \frac{\partial }{\partial \bar{x}}\left(k_{x x} \frac{1}{x^{*}} \frac{\partial \bar{T}}{\partial \bar{x}}\right)+\frac{1}{y^{*}} \frac{\partial }{\partial \bar{y}}\left(k_{y y} \frac{1}{y^{*}} \frac{\partial \bar{T}}{\partial \bar{y}}\right)+\dot{Q}, \quad \textit{while} \quad\dot{Q}=-Q_r \frac{1}{t^{*}} \frac{\partial \bar{\alpha}}{\partial \bar{t}}.
\end{equation}

\begin{figure}[H]
\centerline{\includegraphics[width=1\textwidth]{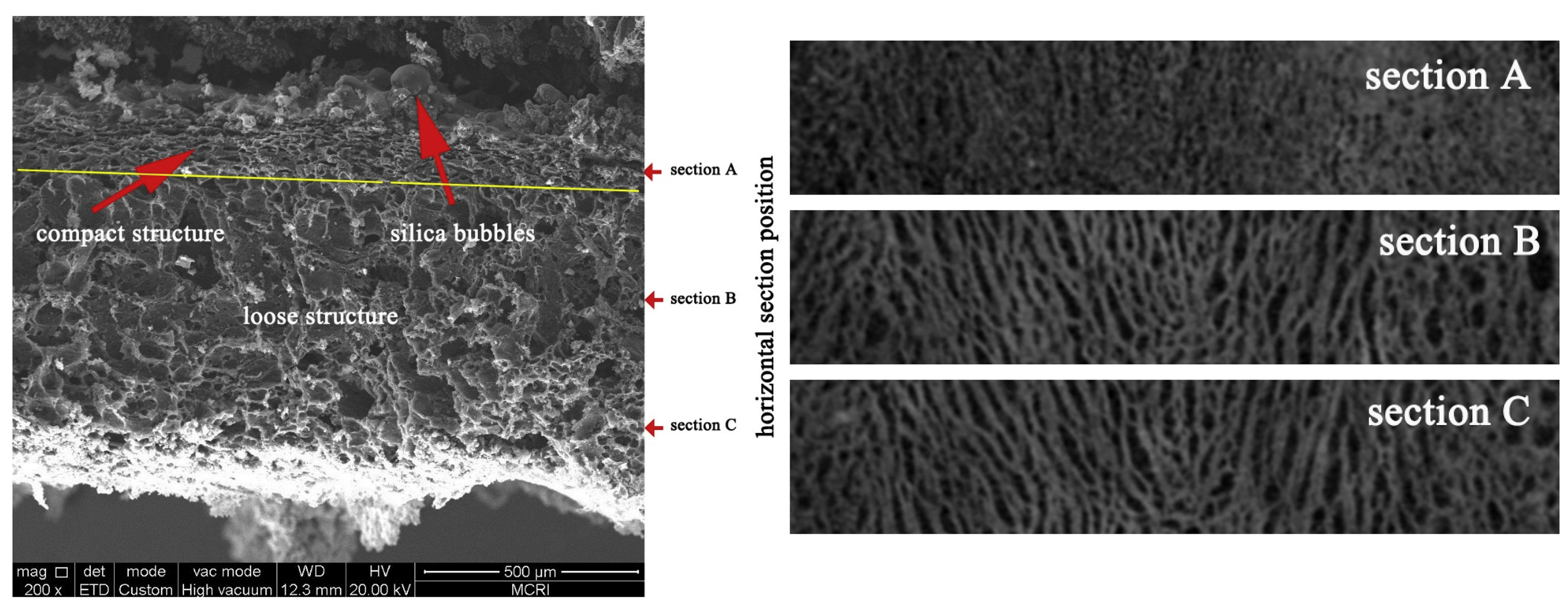}}
\caption{SEM image of vertical-section of the char layer (left) and Micro-CT images of horizontal-section of the char layer (right) \cite{xi2019redeposition}.}
\label{SEM}
\end{figure}

The following steps were required to expand the PINN to 2D: adding one input (y dimension) and providing a pre-layer for the y input, adjusting error calculation to include a term for the second derivative of the PINN prediction with respect to y, and modifying the error calculation for the heat transfer boundary conditions. The total loss function's many terms, as well as their complexity, present huge complications in network training. The only way to get meaningful results from the Adam optimizer was to utilize scored-adaptive weights; however, we have found that the process has a high computational cost and lacks robustness. We offer a unique strategy, inspired by \cite{pirk2017interactive}, to overcome the limitations of standard network training methods for multiphysics issues. 

\begin{figure}[H]
\centerline{\includegraphics[width=.8\textwidth]{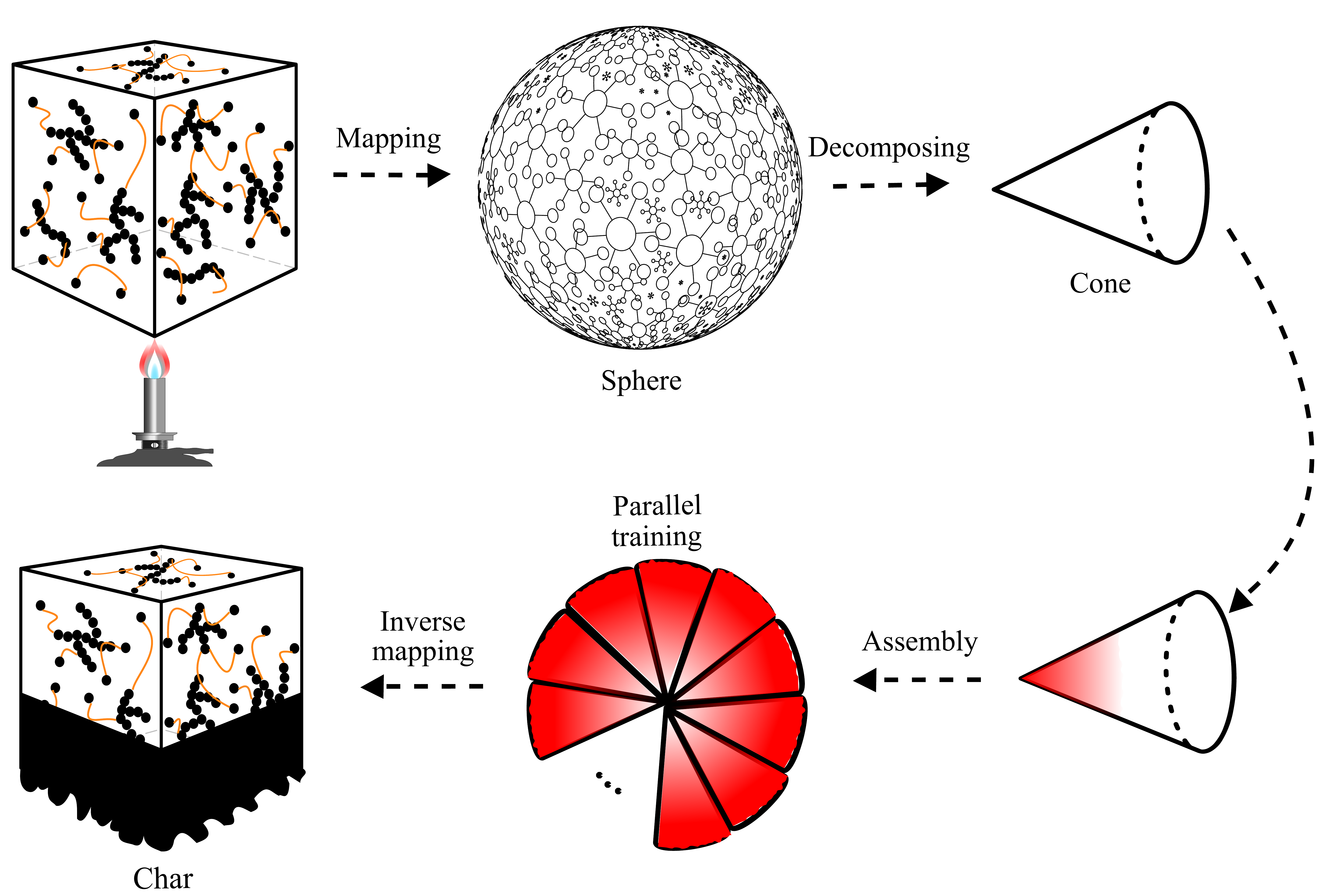}}
\caption{A mesh is used to discretize the material (left), the pyrolyzing front proceeds towards the last spot of burning as pyrolysis progresses (right).}
\label{meshing}
\end{figure}

\begin{figure}[H]
\centerline{\includegraphics[width=.8\textwidth]{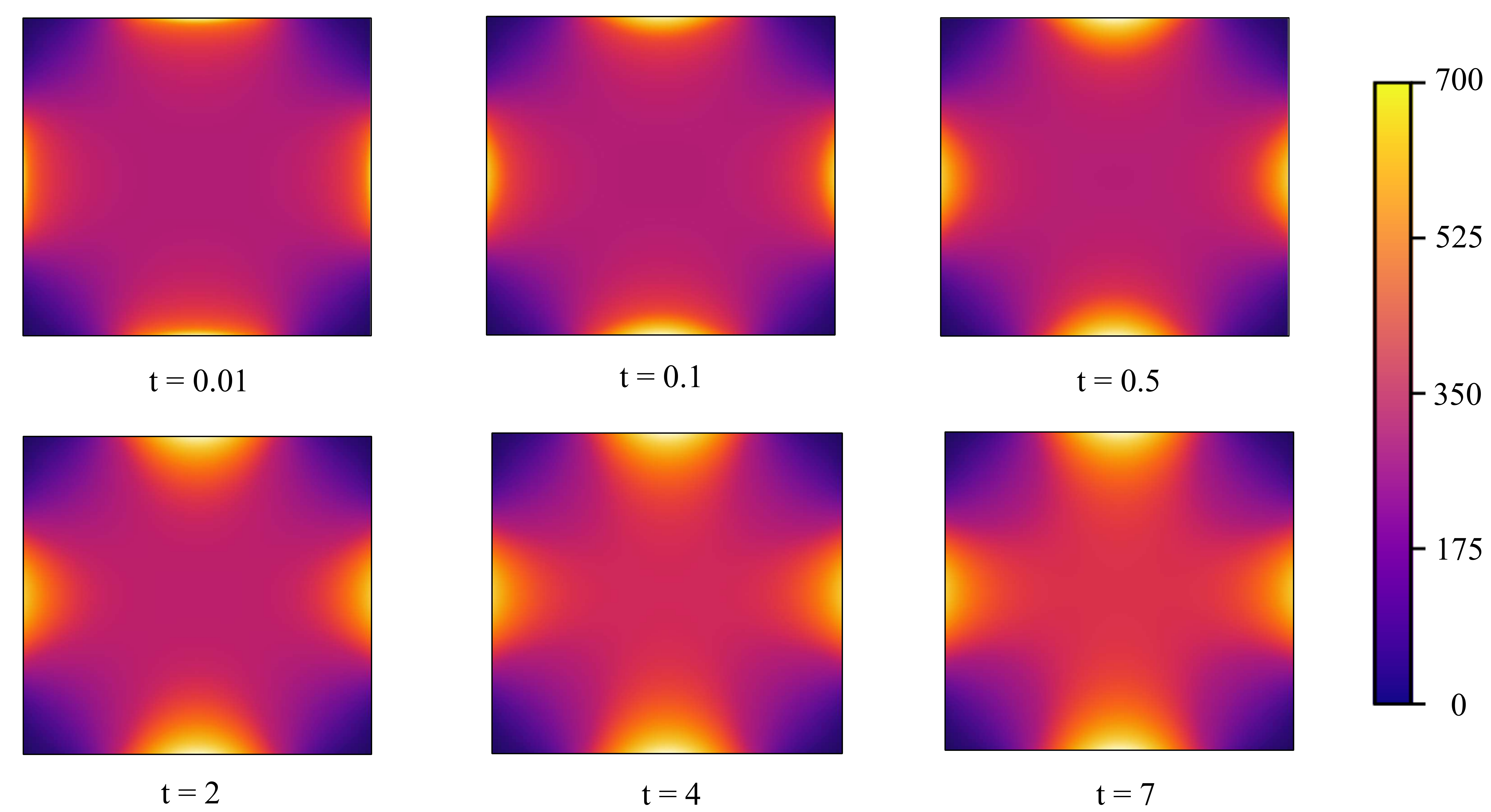}}
\caption{Temperature predictions using PINN for various time steps in burning $t = 0.01, 0.1, 0.5, 2, 4, 7$.}
\label{2Dtemp}
\end{figure}

\begin{figure}[H]
\centerline{\includegraphics[width=.8\textwidth]{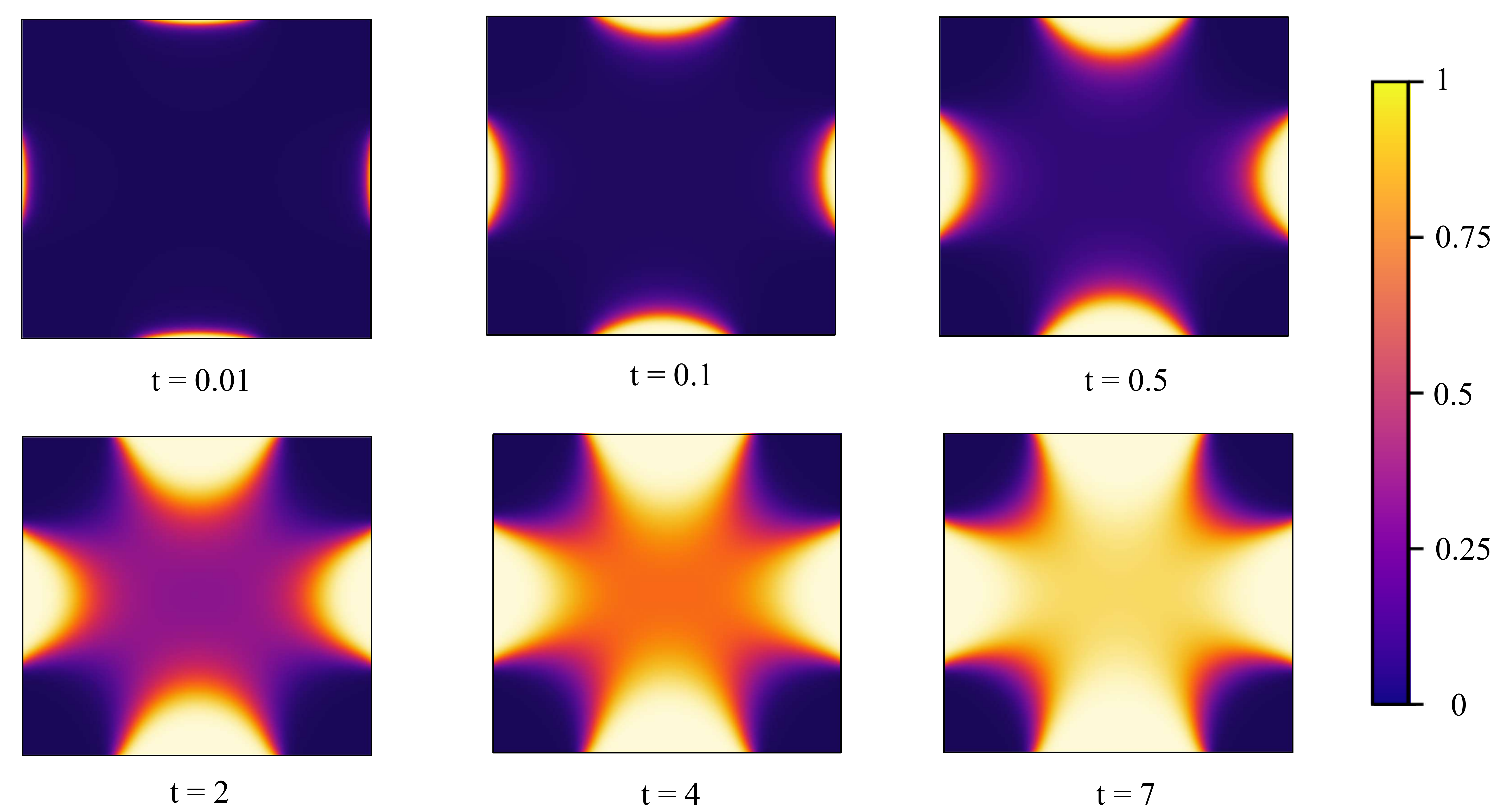}}
\caption{Degree of burning predictions using PINN for various time steps in burning $t = 0.01, 0.1, 0.5, 2, 4, 7$.}
\label{2Dburn}
\end{figure}

The pyrolyzing front is shown as a two-dimensional triangle element. It means that we reduce the order of 2D problems by decomposing them into many 1D problems and training them as a collection of learning. In this concept, we need to assume the last point of burning in the material. Next, we decompose 2D geometry into many 1D elements, which the last point of burning is shared among all 1D elements. Fig. \ref{2Dtemp} and Fig. \ref{2Dburn} show the PINN solution for temperature and degree of burning for different time steps, respectively. The domain is subjected to a constant temperature $300 + 700e^{-15x^2}$ at the top and bottom faces and a constant temperature $300 + 600e^{-15y^2}$ at the right and left faces.

\section{Case Study 3:  Shen's Model for Burning}
\label{shen}
The second category of burning, which considers that the specific heat remains constant, with only thermal conductivity and material density being temperature dependent, is presented in this section \cite{shen2007modeling}. The heat transfer PDE in a polymeric material is presented as

\begin{equation}
\label{pde5}
    \frac{\partial}{\partial t}\left(\rho_s C_{s} T + \rho_c C_{c} T\right)=\frac{\partial}{\partial x}\left(k_{x x} \frac{\partial T}{\partial x}\right)+\dot{Q}, \quad \text{while} \quad \dot{Q} = K\rho_s (C_s - C_c) (T).
\end{equation}

\begin{figure}[H]
\centerline{\includegraphics[width=.5\textwidth]{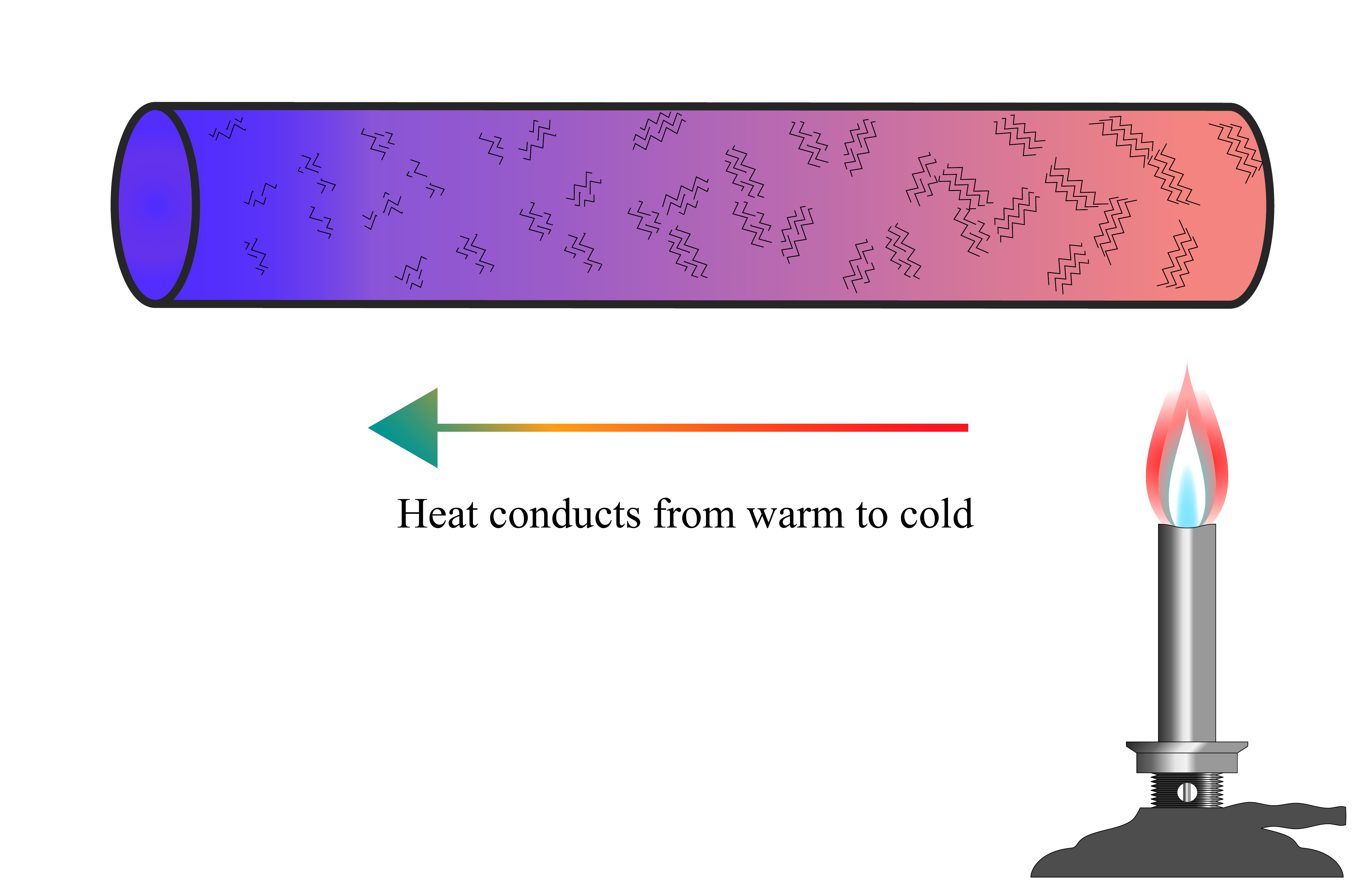}}
\caption{Schematic of 1D burning and char formation.}
\label{conduct}
\end{figure}

Results are presented for boundary and initial conditions such that $L = 1 mm$, $T_{0} = 700x^3$, $T_b = 0$ and $T_t = 700$ for duration of $10$ seconds. The training is conducted on $500$ uniformly distributed points in $(x,t)$. 

\textbf{Thermal characteristic values of polymers}

\begin{figure}[H]
\centerline{\includegraphics[width=.9\textwidth]{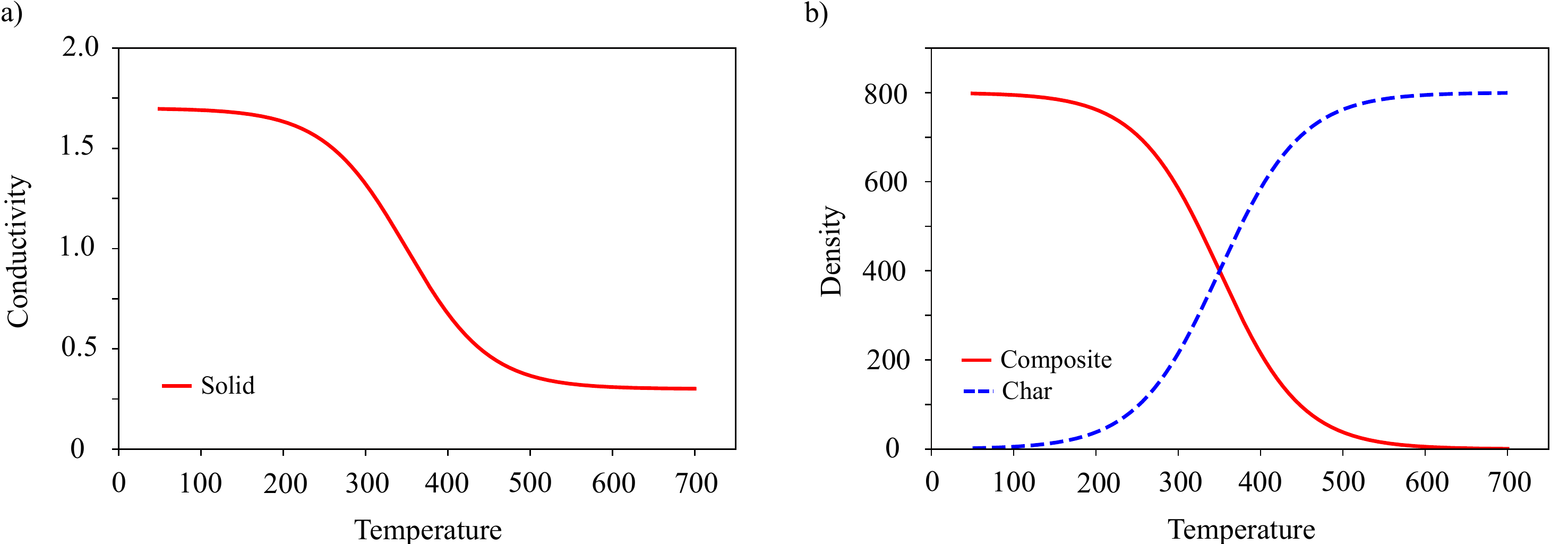}}
\caption{Thermo-physical properties of polymer}
\label{cond+dens}
\end{figure}

The thermal conductivity and density of the polymer, as well as char, have temperature-dependent properties, as shown in Fig. \ref{cond+dens}. The thermal conductivity of the material is shown in Fig. \ref{cond+dens}.a. When the temperature rises above $250 F^{\circ}$, the sample's exposure time to heat flux increases, and the material's thermal conductivity rapidly decreases as a result of the temperature rise. The material's thermal conductivity before this temperature is around $10 W/m K$. The thermal conductivity of the solid stabilizes at $0.3 W/m K$ at temperatures up to $550 F^{\circ}$. Beyond $250$ degrees Fahrenheit, the density of the polymer rapidly declines until it reaches zero at $550$ degrees Fahrenheit. The density of the char, on the other hand, rapidly increases to $800 kg/m3$ at temperature $700$ (see Fig. \ref{cond+dens}.b).

\textbf{Comparison between numerical and predicted temperature profiles}

\begin{figure}[H]
\centerline{\includegraphics[width=.9\textwidth]{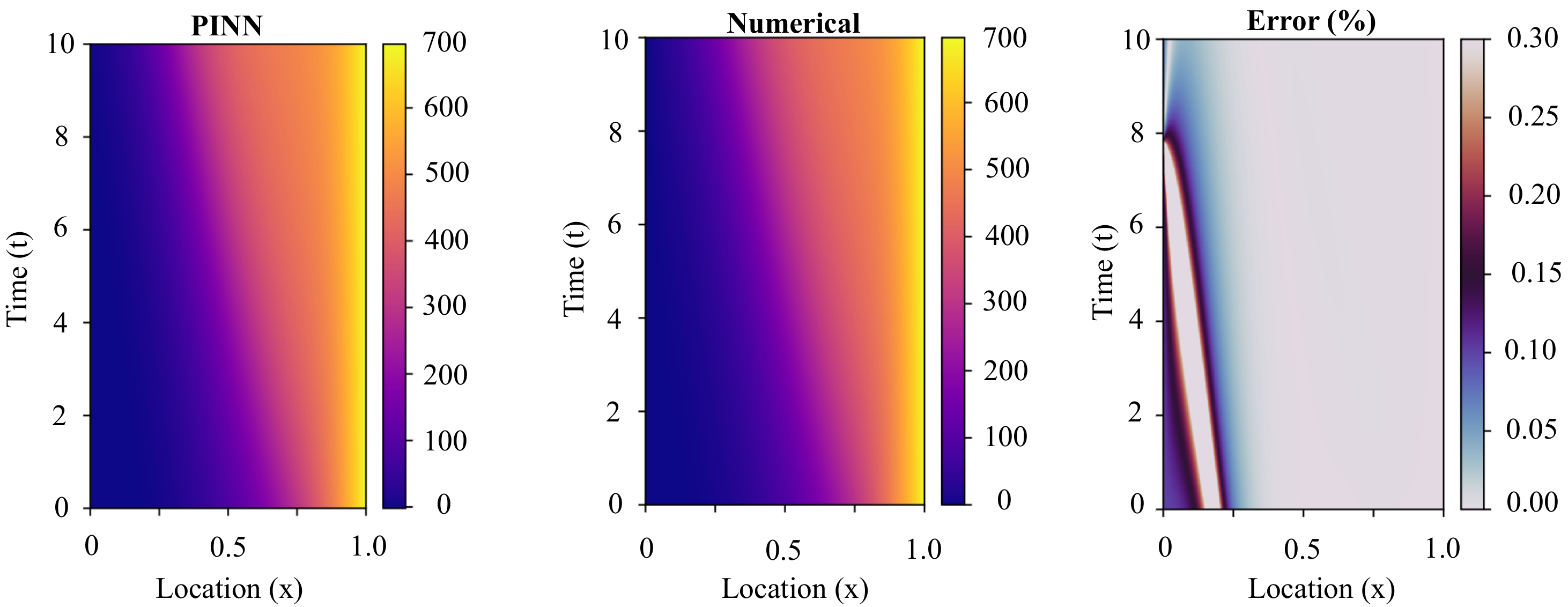}}
    \caption{PINN and numerical temperature predictions for full cycle of burning.}
    \label{res}
\end{figure}

To assess the accuracy of the PINN in terms of predicting $T$, results are compared to numerical predictions from MATLAB. Fig. \ref{res} shows the temperature prediction for the case study. The obtained relative error of the predicted temperature when comparing PINN and numerical methods is below
$1 \%$. 

\begin{figure}[H]
\centerline{\includegraphics[width=.9\textwidth]{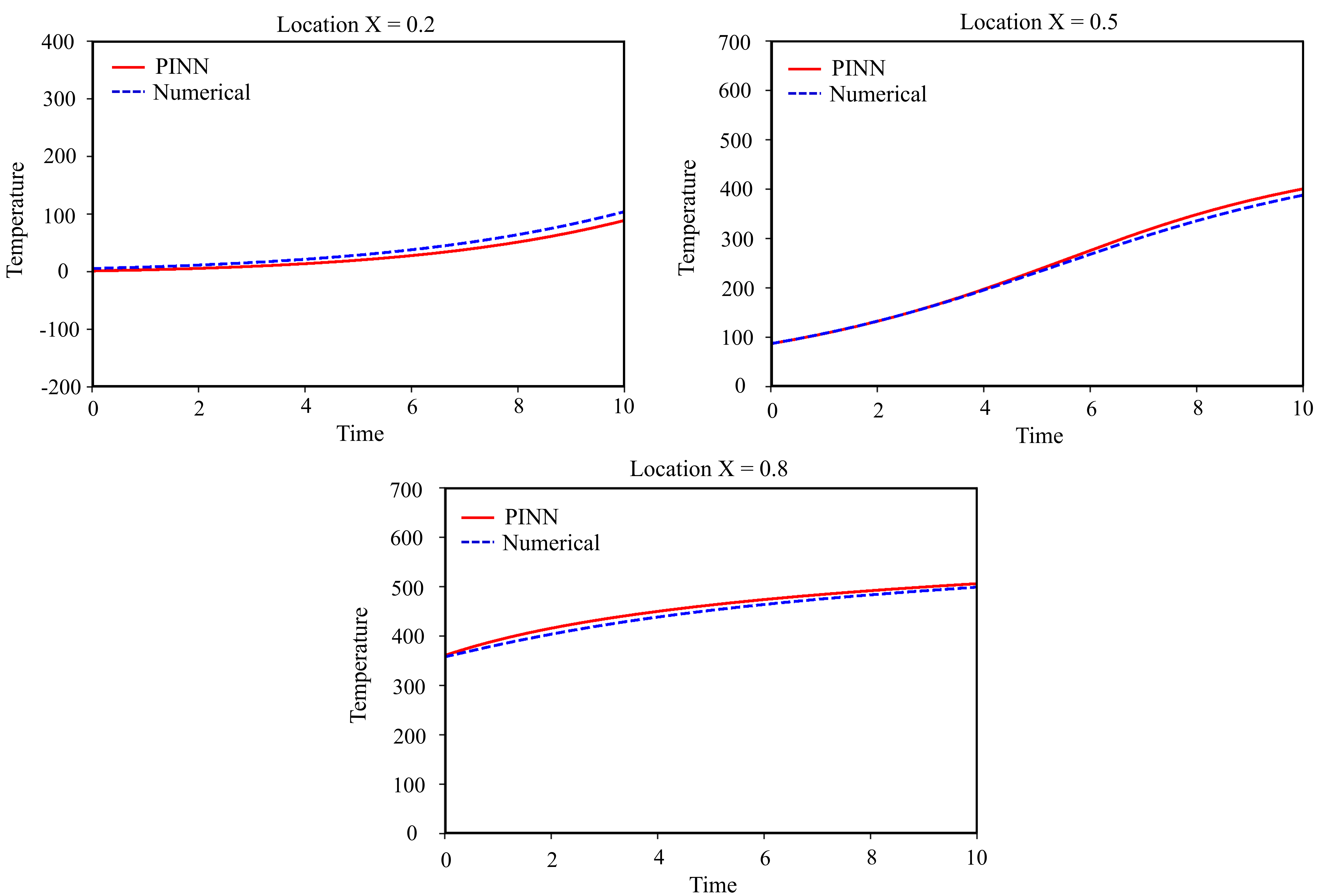}}
    \caption{Temperature history for (x = 0.2), middle of polymer (x = 0.5), and for boundary (x = 0.8).}
    \label{location}
\end{figure}

\begin{figure}[H]
\centerline{\includegraphics[width=.6\textwidth]{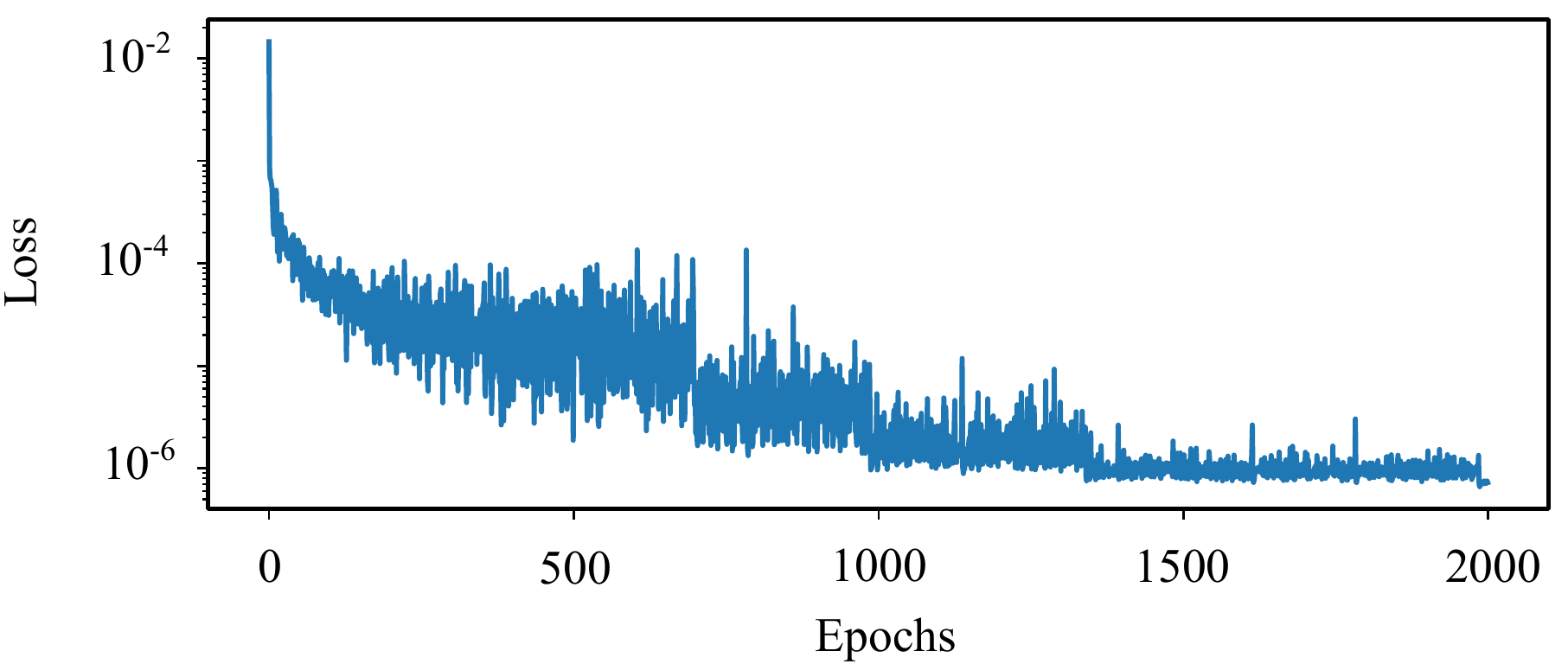}}
    \caption{Evolution of loss term in the training of the network.}
    \label{loss}
\end{figure}

The temperature evolution at the left side of the polymeric material, middle of the polymeric material, and right side of the polymer are shown for the full burning cycle in Fig. \ref{location}. Also, The mean square error of the loss function is reported in Fig. \ref{loss}.

It is somewhat remarkable that a neural network with hundreds of parameters cannot approximate the analytical solution of the 1D hyperbolic PDE with a non-convex flux function with any degree of accuracy. This is particularly striking because, in theory, there should exist a network that can offer a near approximation of the continuous solution for any arbitrarily chosen PDE. However, this is not what is seen.
As a result, this leads us to conclude that the issue is not with the solution in nature but with the methodology we employ to obtain it, i.e., the optimization procedure or the loss function.

\section{Discussion and Concluding Remarks}
\label{con}
In this paper, we studied a comprehensive study on Physics-Informed Neural Networks (PINNs) for the forward solution of  pyrolysis problems by making the training more straightforward. We explored how Physics-Informed Neural Networks (PINNs) can be employed to solve pyrolysis problems in the forward phase. Our research is the first to explore fully coupled temperature-degree-of-burning relationships in pyrolysis problems. We presented a dimensionless version of these relations that leads to the optimizer's stable and convergent behavior.

While the achieved results are close to our expectations, it should be noted that training PINNs is time-consuming. We relate the training challenge to the multi-objective optimization issue and the application of a first-order optimization algorithm, as reported by others. Given the difficulties encountered and overcome in this work for the forward problem, the next step is to use PINNs to inverse burning situations.


A multi-task learning approach has emerged in which a NN must fit observed data while decreasing a PDE residual. This article introduces single- and separated-network PINN architectures to forecast temperature distributions and the degree of burning of a pyrolysis problem in a one-dimensional (1D) and two-dimensional (2D) rectangular domain for the first time to model through collocation training.

The complex and non-convex multi-objective loss function presents substantial obstacles for forward problems in training PINNs. We discovered that adding several differential relations to the loss function causes an unstable optimization issue, which may lead to convergence to the trivial null solution or significant deviation of the solution.
The dimensionless form of the coupled governing equations that we find most beneficial to the optimizer is used to address this problem. The numerical results are compared with results obtained from PINN to show the performance of the solution. Our research is the first to explore fully coupled temperature-degree-of-burning relationships in pyrolysis problems. Unlike classical numerical methods, the proposed PINN does not depend on domain discretization. In addition to these characteristics, the proposed PINN achieves good accuracy in predicting solution variables, which makes it a candidate to be utilized for surrogate modeling of pyrolysis problems.
While the achieved results are close to our expectations, it should be noted that training PINNs is time-consuming.

\appendix

\pagebreak

\appendix
\section{General PDEs}

In the case of a function $f: \mathbb{R}^{n} \rightarrow \mathbb{R}^{m}$ for which we want to calculate the Jacobian $J$, AD can calculate the numerical derivative in either forward or reverse mode once all operations' graphs that make up the mathematical expression are calculated. As a result, PINNs can solve differential equations in their most general form, such as

\begin{equation}
\begin{aligned}
\mathcal{F}(\boldsymbol{u}(\boldsymbol{z}) ; \gamma) &=\boldsymbol{f}(\boldsymbol{z}) & & \boldsymbol{z} \text { in } \Omega, \\
\mathcal{B}(\boldsymbol{u}(\boldsymbol{z})) &=\boldsymbol{g}(\boldsymbol{z}) & & \boldsymbol{z} \text { in } \partial \Omega,
\end{aligned}
\end{equation}
defined with the boundary $\partial \Omega$ on the domain $\Omega$. $\boldsymbol{z}=\left[x_{1}, \ldots, x_{n} ; t\right]$ denotes the space-time coordinate vector, $\boldsymbol{u}$ denotes the unknown solution, $\gamma$ are the physics parameters, $\boldsymbol{f}$ denotes the function identifying the problem's data, and $\mathcal{F}$ denotes the non linear differential operator. Finally, because the initial condition is a sort of Dirichlet boundary condition in the spatio-temporal domain, $\mathcal{B}$ can be used to signify arbitrary initial or boundary conditions relating to the problem, and $\boldsymbol{g}$ can be used to denote the boundary function.

In essence, AD includes a PDE into the loss of the neural network, where the differential equation residual is

\begin{equation}
r_{\Theta}(\boldsymbol{z}):=\mathcal{F}\left(\hat{\boldsymbol{u}}_{\Theta}(\boldsymbol{z}) ; \gamma\right)-\boldsymbol{f}.
\end{equation}

The following is the result of the original formulation of the aforementioned equation.

\begin{equation}
r_{\Theta}(\boldsymbol{x}, t):=\frac{\partial}{\partial t} \hat{u}_{\Theta}(\boldsymbol{x}, t)+\mathcal{F}_{\boldsymbol{x}} \hat{u}_{\Theta}(\boldsymbol{x}, t).
\end{equation}

In the deep learning framework, the chain rule is employed to calculate hierarchical derivatives from the output layer to the input layer with regard to input spatio-temporal coordinates using automatic differentiation for numerous layers.

\section{Mapping}
We'll discuss how to map a circular region to a square region and back in this part. There are countless variations on how to carry out this mapping. We focus on mappings that have attractive closed-form invertible equations. In terms of mathematics, we are looking for functions $f$ that transform each point $(u, v)$ in the circular disc to a point $(x, y)$ in the square region and vice versa. To put it another way, we're trying to come up with equations for $f$ such that $(u,v) = f(x,y)$ and $(x, y) = f^{-1} (u, v)$. Additionally, we will subject the mapping to the following three constraints:

\begin{itemize}
    \item $(0,0) = f(0,0)$, meaning that the center of the circle and the center of the square, respectively.
    \item The extreme points of the forms all along x-axis match up, or $(\pm 1,0) = f(\pm 1,0)$.
    \item The extreme points of the shapes on the y-axis correspond because $(0, \pm 1) = f(0, \pm 1)$.
\end{itemize}

One of the easiest methods for mapping a circular disc to a square area is to linearly extend the circle to its rim. Depending on which side of the square the stretching takes place, the equations for extending from rim to rim must be divided into four separate situations. We may further reduce these equations by applying the signum function, and a technique developed by Dave Cline \cite{shirley1997low} to get:

\begin{equation}
u=\left\{\begin{array}{ll}
\operatorname{sgn}(x) \frac{x^{2}}{\sqrt{x^{2}+y^{2}}} & \text { when } x^{2} \geq y^{2} \\
\operatorname{sgn}(y) \frac{x y}{\sqrt{x^{2}+y^{2}}} & \text { when } x^{2}<y^{2}
\end{array} \quad v= \begin{cases}\operatorname{sgn}(x) \frac{x y}{\sqrt{x^{2}+y^{2}}} & \text { when } x^{2} \geq y^{2} \\
\operatorname{sgn}(y) \frac{y^{2}}{\sqrt{x^{2}+y^{2}}} & \text { when } x^{2}<y^{2}\end{cases}\right.
\end{equation}

Also, Manuel Fernandez Guasti developed an algebraic equation in 1992 to depict a form that is between a circle and a square \cite{guasti1992analytic}. The parameter $s$ in his equation may be utilized to nicely mix the circle and the square. We can create a method to seamlessly transfer a circular disc to a square area using the Fernandez Guasti squircle.
The key concept is to map each circular contour on the inside of the disc to a square circular contour. This may be combined with the radial restriction, and the resulting equations for the mapping can be obtained.

\begin{figure}[H]
\centerline{\includegraphics[width=.9\textwidth]{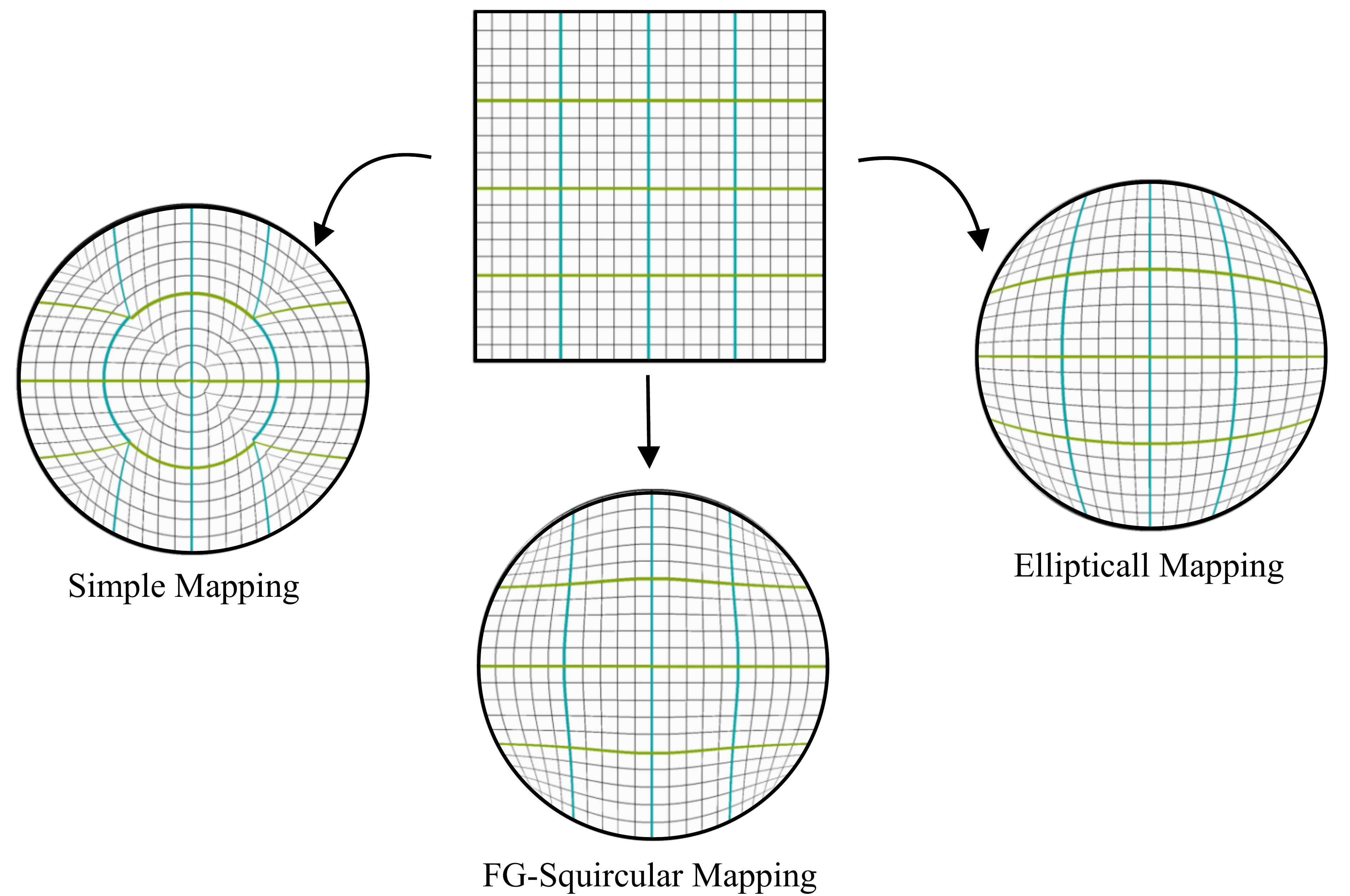}}
    \caption{Various type of mapping.}
    \label{mapping}
\end{figure}

On the other hand, a regular rectangular grid is transformed into a regular curvilinear grid made up of elliptical arcs via elliptical grid mapping. The inverse of Nowell's elliptical grid mapping is now obtained. The three steps of the derivation can be summed up as follows:
Step 1: Introduce the trigonometric variables $\alpha$ and $\beta$ and convert circular coordinates $(u, v)$ to polar coordinates.
Step2: Obtain an equation for $(x,y)$ in terms of $\alpha$ and $\beta$ in step two.
Find an equation for $(x,y)$ in terms of $u$ and $v$ in step three.

\bibliographystyle{main}
\bibliography{Polymer.bib}

\end{document}